\documentclass{svjour3}                     
\smartqed  
\usepackage{graphicx}
\usepackage{threeparttable}
\usepackage{mathptmx}      
\usepackage{natbib}
\usepackage{color}
\usepackage{url}
\usepackage{textcomp}
\usepackage{fixltx2e}
\newcommand\ion[2]{#1$\;${\scshape{#2}}}
\begin{document}
\title{Accurate abundance analysis of late-type stars: advances in atomic physics}
\subtitle{}


\author{Paul S. Barklem}


\institute{P. S. Barklem \at
              Theoretical Astrophysics, Department of Physics and Astronomy, Uppsala University, \\ 
              Box 516, SE-751 20 Uppsala, Sweden \\
              Tel.: +46-18-4715981\\
              \email{paul.barklem@physics.uu.se}          
              }

\date{Received: date / Accepted: date}

\maketitle

\begin{abstract}
The measurement of stellar properties such as chemical compositions, masses and ages, through stellar spectra, is a fundamental problem in astrophysics.  Progress in the understanding, calculation and measurement of atomic properties and processes relevant to the high-accuracy analysis of F-, G-, and K-type stellar spectra is reviewed, with particular emphasis on abundance analysis.  This includes fundamental atomic data such as energy levels, wavelengths, and transition probabilities, as well as processes of photoionisation, collisional broadening and inelastic collisions.  A recurring theme throughout the review is the interplay between theoretical atomic physics, laboratory measurements, and astrophysical modelling, all of which contribute to our understanding of atoms and atomic processes, as well as to modelling stellar spectra.  
\end{abstract}

\vspace*{5mm}
\noindent
{\bf Abbreviations}

\vspace*{3mm}
\noindent
\begin{tabular}{lp{8.5cm}}
ABO  & Anstee, Barklem \& O'Mara \\
ASD  & Atomic Spectra Database (at NIST) \\
ATLAS & Model stellar atmosphere computer program by Kurucz \\
AUTOSTRUCTURE & Atomic structure computer program by Badnell \\
BSR & $B$-Spline $R$-matrix  \\
CCC  & Convergent Close Coupling \\
CCD  & Charge-Coupled Device \\
CIV3 & Atomic structure computer program by Hibbert \\
DESIREE & Double ElectroStatic Ion Ring ExpEriment \\
DSB   & Derouich, Sahal-Br\'echot \& Barklem \\
ESO   & European Southern Observatory \\
FARM   & $R$-matrix computer program for external region problem by Burke \& Noble \\
HBOP & Hydrogen Bound and bound-free OPacity code \\
HLINOP & Hydrogen LINe OPacity computer code by Barklem \& Piskunov \\
HLINPROF & Hydrogen LINe PROFile computer code by Barklem \& Piskunov\\
IDL & Interactive Data Language \\
KAULAKYS & Computer program for Kaulaky's free-electron model by Barklem \\
LFU & Lindholm-Foley-Uns\"old \\
LTE  & Local Thermodynamic Equilibrium \\
MARCS & Model stellar atmosphere computer program by Gustafsson et al.\\
MOOG & Stellar spectrum synthesis computer program by Sneden \\
MSWAVEF & Mometum-Space WAVEFunction computer code by Barklem \\
NIST & National Institute of Standards and Technology (U.S.A.) \\
RMATRX I & $R$-matrix computer program for internal region problem by Berrington et al. \\
RMPS & $R$-Matrix with Pseudo-States \\
STARK-B & Stark broadening database \\
STGF   & $R$-matrix computer program for external region problem by Berrington et al. \\
SUPERSTRUCTURE & Atomic structure computer program by Eissner et al. \\
TOPbase &  Opacity Project on-line atomic database \\
VALD & Vienna Atomic Line Database \\
VAMDC & Virtual Atomic and Molecular Data Centre \\
\end{tabular}

\section{Introduction}
\label{intro}

It is just over one hundred years since Niels Bohr proposed his model of the hydrogen atom \citep{phil_i_1913, phil_xxxvii_1913}.  Stellar spectra played a significant role in Bohr's considerations regarding atomic structure \citep[e.g.][]{bohr_spectra_1913}, and in turn paved the way for the study of stars via their spectra.  The quantum revolution born out of the Bohr model, opened the way to understanding atomic structure and the wide variety of radiative and collision processes that shape the spectra of stars.  However, despite the amazing success of quantum theory, our understanding of the atomic processes in stellar atmospheres is by no means a closed subject.  The myriad of processes occurring in stellar atmospheres and the difficulty in applying quantum mechanics to many-body problems (the scattering of two electrons by a proton is still an active field of research) means there is there still some way to go before we have a quantitative understanding of these processes at the accuracies we would aspire to for the analysis of stellar spectra.

\begin{figure}
\center
\includegraphics[width=0.75\textwidth]{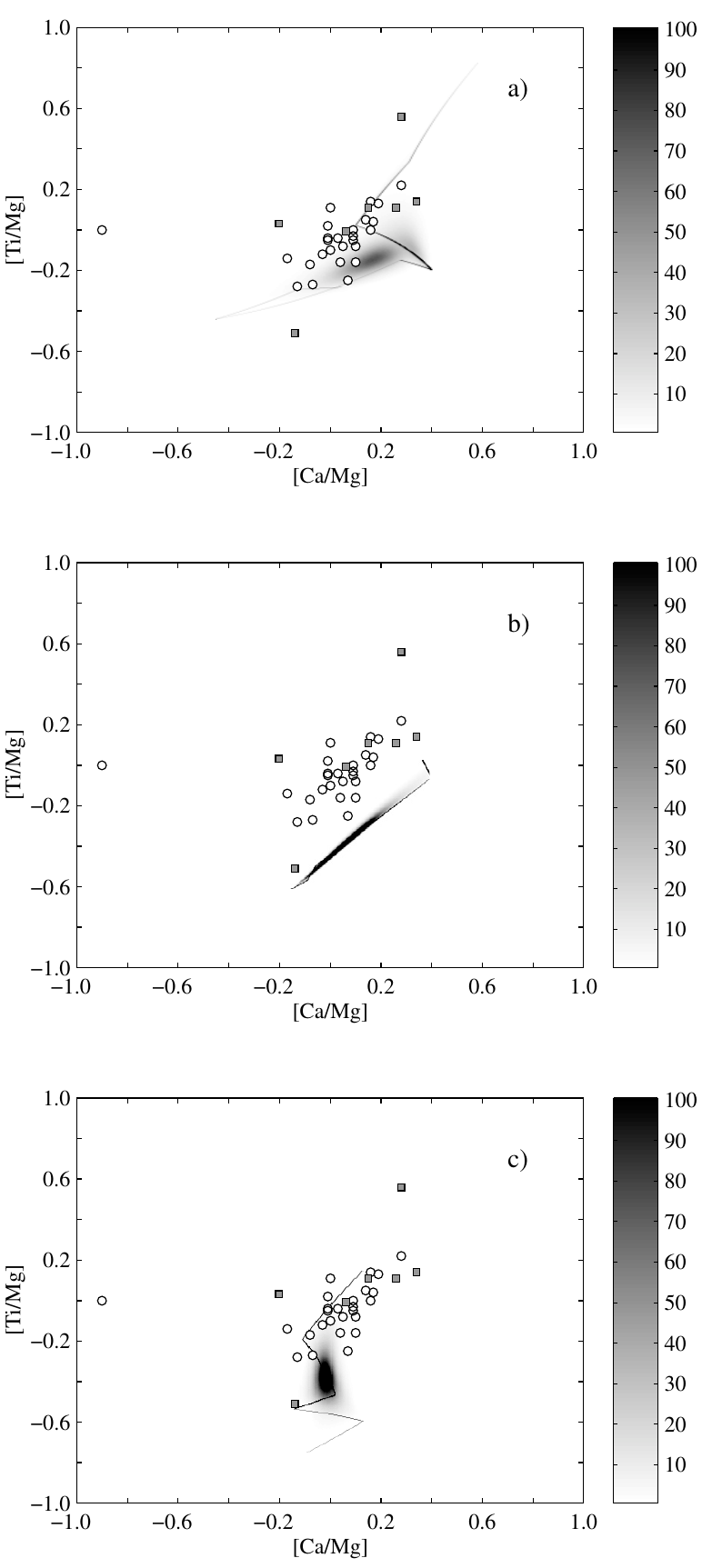}
\caption{Distribution of low-mass stars in the [Ti/Mg]--[Ca/Mg] plane. Small circles denote the observations of extremely metal-poor giants ([Mg/H] $< −2.5$) by \cite{cayrel_first_2004}, and the gray squares denote a corresponding sample ([Mg/H] $< −2.5$) of dwarfs by \cite{cohen_abundances_2004}. Theoretical distributions are shown using yields by a) \cite{woosley_evolution_1995}, b) \cite{umeda_nucleosynthesis_2002}, and c)  \cite{chieffi_explosive_2004}. Credit: \citeauthor{Karlsson2005a}, A\&A, 436, 879, 2005, reproduced with permission \textcopyright\ ESO.}
\label{fig:karlsson_sss}     
\end{figure}

The measurement of stellar properties such as chemical compositions, masses and ages, through stellar spectra, is a fundamental problem in astrophysics.  While stellar spectral lines are now routinely measured at accuracies of $<0.01$~dex (of order 1\%) or better, systematic errors often dominate the interpretation such that the actual errors in derived properties, for example in the abundance of a chemical element, are at least an order of magnitude larger.  It is completely reasonable to expect that nature has more information to reveal if we could take the step to the next level and derive abundances with relative accuracies\footnote{It is important to note the distinction between accuracy and precision.  Accuracy refers to the correctness of a measurement, i.e. how close the measurement is to the true value, while precision refers to the reproducibility of a measurement.  Depending on the situation, \emph{absolute} or \emph{relative} accuracy may be most important.} approaching 0.01~dex.  For example, \cite{Karlsson2005a} have theoretically predicted structures in abundance ratios that may be the result of stars enriched by a single supernova event (``Single Supernova Sequence'') and could thus potentially probe nucleosynthesis at early times in the evolution of the Galaxy.  As an example, fig. 11 of \citeauthor{Karlsson2005a} is reproduced in fig.~\ref{fig:karlsson_sss} (see also figs.~9 and 16--18 of that paper).  It demonstrates, first that such structures could potentially be used to distinguish various supernova models, and second that small samples of stars with abundance ratios with relative accuracy of order 0.1~dex (25\%) are not sufficient to resolve them. They conclude that detecting such structures will require samples of hundreds of metal-poor stars, stars with metal abundances significantly lower than the solar value and thus probing the evolution of the Galaxy at early times, studied at very high spectroscopic accuracy. \citet{Lindegren2013} have used Monte-Carlo simulations to examine under which conditions structures in abundances can realistically be distinguished. Table~\ref{tab:resolve} gives some representative examples of sample sizes required to resolve equal Gaussian populations separated by a given amount in abundance, for various chosen values of the measurement error.  Note, the measurement errors in such a situation may arise from systematic sources in interpretation of abundances, e.g. the lack of relative accuracy due to incorrect account for differences in physics between stars with different properties, or due to lack of precision in the measurement of the spectra, e.g. due to noise. The results of \citeauthor{Lindegren2013} and those examples in Table~\ref{tab:resolve} clearly show that large-number statistics cannot fully compensate for errors in the abundance measurements, and that a rather modest improvement in measurement errors has a powerful effect.

%
\begin{table}
\caption{Approximate minimum sample sizes needed to resolve populations separated by 0.2, 0.1, and 0.01~dex in abundances, for measurement errors of 0.1, 0.05, and 0.01 dex, as derived by \citet{Lindegren2013}.}
\label{tab:resolve}       
\begin{tabular}{rrrr}
\hline\noalign{\smallskip}
Measurement    & \multicolumn{3}{c}{Resolve populations separated by:}  \\
error          &  0.2~dex & 0.1~dex & 0.01~dex  \\
\noalign{\smallskip}\hline\noalign{\smallskip}
0.1 dex (25\%)  & $\sim 3000$ & $> 10^5$    & $\gg 10^5$  \\
0.05 dex (12\%) & $\sim 100$  & $\sim 3000$ & $\gg 10^5$  \\
0.01 dex (2\%)  & $\ll 30$    & $< 30$      & $> 10^5$    \\
\noalign{\smallskip}\hline
\end{tabular}
\end{table}

Advances in the accuracy of stellar abundances are needed to gain insight into many current problems in modern astrophysics.  The above case of Single Supernovae Sequences is a specific example of what is often referred to as ``galactic archeology'': the reconstruction of the history of the Milky Way via the information encoded in the dynamics and chemical abundances of present-day, long-lived stars.  Other problems include the understanding of the role of chemistry in the formation of planetary systems such as our own, problems in stellar physics, and in cosmology \citep[e.g.][]{melendez_solar_2010a,Gustafsson2010,tuccimaia_high_2014}.  It again seems reasonable that more accurate abundances will shed light on these problems.

The focus of this review is late-type stars, specifically F-, G- and K-type stars.   These stars span the regime between significant molecule formation at the cool end, and where hydrogen and all other elements become ionised at the hot end.  At the cool end, molecules begin to dominate the visual spectrum making analysis highly uncertain due to the paucity of clean spectral lines.  At the hot end, stellar rotation and the lack of lines of ionised species arising in the visual part of the spectrum hinders chemical abundance analysis; in addition stellar lifetimes are shorter and thus hotter stars are not as useful as temporal probes, e.g. in galactic archeology.   For these reasons, FGK stars are the classical tracers of chemical composition and evolution in the local universe.

The vast majority of analyses of FGK stellar spectra are done with the classical 1D-LTE technique; i.e. one-dimensional (1D), hydrostatic model atmospheres and the assumption of local thermodynamic equilibrium (LTE).  The assumptions of LTE and horizontal homogeneity, leading to need for the approximate treatment of convection (including the microturbulence parameter), together with errors in the underlying atomic data, are presently considered the main sources of systematic errors in abundance analyses, at least as regards theoretical modelling\footnote{Note, errors on the observational side, such as issues in the reduction of echelle spectra, not least continuum placement, may be at least equally important in specific cases \citep[e.g.][]{caffau_photospheric_2008}.}.  These errors affect not only the modelling of the lines of interest directly, but also usually indirectly through the stellar parameters derived from other lines.  Realistic treatment of radiation and convection is done through non-local thermodynamic equilibrium (non-LTE) radiative transfer and 3D hydrodynamical model atmospheres, and are the subject of intense study \citep[e.g.][]{Asplund2005}.  This review focusses on the atomic physics aspect, in particular the current status of atomic data needed for modelling the photospheres of FGK stars and their spectra.  The photospheres of these stars are dominated by neutral hydrogen, together with the remaining elements in neutral or singly ionised form.  Thus, to model the radiative energy transfer and the spectra of such stars we must understand the structure of relevant atoms and ions, as well as the radiative processes they undergo.  We must also understand the effects of the environment on the atoms and ions, in particular the effects of nearby particles via collisional processes.

\section{Atomic structure: energy levels, wavelengths and transition probabilities}
\label{sec:structure}

Knowledge of atomic structure is, of course, fundamental to interpretation of stellar spectra.  To be precise, it is not usually a complete description of the atomic structure (i.e. the electronic wavefunctions themselves) that is needed, but information related to the observable products of that structure, in particular energy levels and their classifications, wavelengths and transition probabilities.  Ostensibly, LTE requires this data only for lines of interest; however it should be noted that energy-level data for the entire atom are required to calculate the partition function.  Non-LTE requires all these data, namely energy levels, wavelengths and transition probabilities, as well as various other radiative and collisional data, for the entire atom.  Further, that data are needed only for the atom of interest first assumes that the lines of interest are clean and unblended, which is practically never true in stellar spectra (or at least it is difficult to be certain).  Moreover, analysis usually requires theoretical model atmospheres, the calculation of which depends on knowledge of the entire spectrum.  Thus we need information on the spectra of all species of possible importance.  The importance of such data cannot be overstated: it is at the basis of our ability to model the structure of stellar atmospheres and of our ability to make informed choices of lines for measurements of abundances. Yet, despite the over one hundred years since the Bohr model and the inception of quantum mechanics, we are still some way from having the understanding of atoms and their spectra in terms of both the completeness and accuracy that are desired for accurate and \emph{certain} interpretation of stellar spectra and measurement of chemical abundances.

Unfortunately, the calculation of energy levels, wavelengths and transition probabilities is generally not presently possible with the accuracy required for analysis of high-resolution stellar spectra.  For example, \citet[][see tables 2 and 3]{FroeseFischer2009} compares errors in transition energies for \ion{S}{i} and \ion{Ar}{i} from two types of modern atomic structure calculations, and finds errors of order of a per cent, which translates to errors of order 10~\AA\ in the UV, and 100~\AA\ in the infrared.  Obviously, this is not sufficient for predicting the positions of spectral lines in stellar spectra, where lines have typical widths and separations of less than 1~\AA. Therefore laboratory measurements must be done to obtain energy levels and wavelengths with the desired accuracies.  We note, however, the very important role of theoretical calculations in classifying the levels: to pick a relatively recent example, see \cite{oliver_energy_2008a} for a study of \ion{Sn}{i}.  For transition probabilities the differences are less stark:  \cite[][see tables 2 and 3, and fig.~1]{FroeseFischer2009} shows that errors in theoretical transition probabilities are typically estimated to be of order 10-20\% \citep[see, e.g. also][for a recent example]{malcheva_radiative_2015}, but can be much larger, especially for weak lines.  Note also that these results are for relatively simple atoms and that complex atoms are significantly more difficult.  Laboratory measurements of transitions probabilities generally have uncertainties of similar order, though more consistently (i.e. the much larger errors appearing in the results of theoretical calculations are not seen), and with complex atoms being accessible \citep[see, e.g.][table 4]{Wood2013}.  Thus, laboratory measurements are again preferable also for transition probabilities; however, theoretical calculations are an important complement for transitions not yet measured in the laboratory.

The NIST Atomic Spectra Database (ASD) \citep{ASD} critically compiles energy levels, wavelength and transition probability data, and as such is an invaluable resource for analysis of stellar spectra.  The database is constantly being improved and updated \citep[see][]{reader_nist_2012,kramida_current_2014}\footnote{The webpage \texttt{http://physics.nist.gov/PhysRefData/ASD/Html/verhist.shtml} logs all improvements and updates.}.  This includes reevaluation of existing data; for example the data for \ion{Mn}{ii} has been recently reevaluated by \cite{kramida_energy_2013}, making use of a recent freely available code to derive optimised energies and Ritz  wavelengths \citep{kramida_program_2011-1}.   A particularly important recent improvement regarding energy levels is the inclusion of a new analysis of \ion{Fe}{ii}, one of the most important ions in astrophysical spectroscopy \citep{nave_spectrum_2013}.  We note that the late S.\ Johansson spent a ``half-life'' studying \ion{Fe}{ii}, giving an indication of the amount of work required for precise laboratory spectroscopy of complex atoms \citep{johansson_halflife_2009-1}. This emphasises the importance of the goal of calculations with spectroscopic accuracy \citep{fischer_calculations_2013,ekman_calculations_2014}, as it would permit large amounts of data to be produced at relatively low cost.  Having said that, laboratory analyses are irreplaceable, and given their usefulness for all kinds of spectroscopy, should be performed.  Regarding transition probabilities, the contributions over recent years of the Wisconsin \citep[e.g.][]{sneden_atomic_2014, lawler_improved_2009}, Imperial College London \citep[e.g.][]{ruffoni_infrared_2013}, Lund-Malm{\"o} \citep[e.g.][]{engstrom_ferrum_2014} and Prague \citep[e.g.][]{civis_infrared_2013} groups have been particularly important.

Due to the considerations above regarding the relative accuracies of theoretical and laboratory data, the energy level data in the ASD is almost exclusively taken from the laboratory, and only lines between states with such data are included.  This, unfortunately does not meet the level of completeness needed for stellar atmosphere modelling and identification of blends in stellar spectra.  In the 1990's \cite{kurucz_why_1990} estimated that roughly half of the intermediate to weak lines in UV of the solar spectrum are missing, and claims, reasonably, that the majority of lines are from the iron group.  Kurucz and co-workers have made significant efforts to supplement laboratory data with semi-empirical data \citep{kurucz_table_1975,Kurucz1992,kurucz_including_2011}, in an attempt to ``include all the lines''.  These calculations make use of Slater-Condon theory via the atomic structure code by \cite{cowan_theoretical_1968,Cowan1981}.  The Slater parameters are determined empirically from observed energy levels, enabling new energy levels and transitions to be predicted, with their transition probabilities calculated by use of scaled Thomas-Fermi-Dirac wavefunctions \citep{Warner1968c}.  The method is detailed in \cite{kurucz_semiempirical_1973,kurucz_semiempirical_1981}.  Considerable progress has been made in these calculations for iron group atoms, with something like a factor 10 increase in the number of lines since the calculations of the 1980's \cite[see tables 3 and 6 of][]{kurucz_including_2011}.  However, Kurucz still estimates that only half the visible lines in the solar spectrum have good wavelengths.  Thus, such semi-empirical calculations are useful for model atmosphere calculations, where the total opacity is relevant, but are less suited to identifying blends and unidentified lines in the spectra.  However, they represent an extremely useful starting point for the identification of more levels and lines in the laboratory.

The above emphasises the natural and necessary interaction between experimental and theoretical studies of atomic structure and atomic spectroscopy.  As mentioned in the introduction, stellar spectra have long been used as a complement to laboratory spectra, as they often provide access to conditions not easily reproducible in the laboratory.  The downside of stellar spectra is naturally that the source shows lines of many atoms, and blending may be an issue.  The availability of extremely high-quality spectral atlases across a very wide spectral range is leading to an increase in the number of analyses making use of both laboratory and stellar spectra.  A particularly successful early example is the identification of highly-excited levels of  \ion{Fe}{i} from joint analysis of laboratory and solar spectra by \cite{nave_highlyexcited_1993,nave_highlyexcited_1993-1}.  They are led to comment that ``All of the newly-identified lines in the [laboratory] spectra correspond to features in the solar spectrum, confirming that the best source of all for \ion{Fe}{i} is the sun''.  More recently, \cite{peterson_new_2015} have identified further  \ion{Fe}{i} levels using a sample of high-quality FGK stellar spectra. \cite{castelli_new_2009} and \cite{castelli_new_2010} have used the early-type chemically-peculiar star, HR~6000, which has a very small $v\sin i$, to identify transitions and levels of \ion{Fe}{ii}.  \cite{castelli_new_2015} have used spectra of the HgMn star HD~175640 to identify energy levels of \ion{Mn}{ii}.  

\begin{figure}
\center
\includegraphics[width=0.95\textwidth]{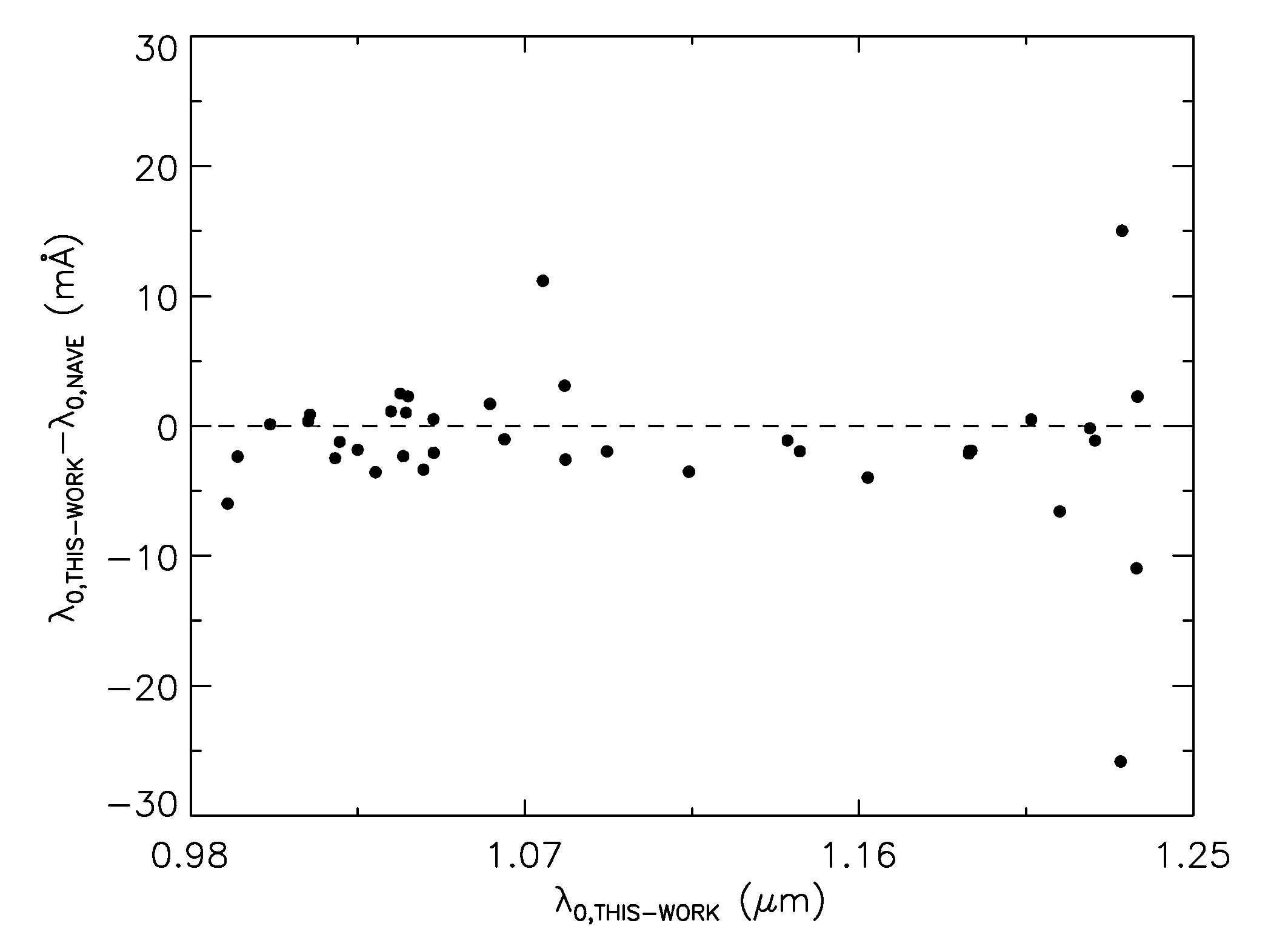}
\caption{Comparison of central wavelengths determined from the solar spectrum and laboratory wavelengths of \cite{Nave1994} for 38 \ion{Fe}{i} lines in the interval 0.98--1.25~$\mu$m. The mean and rms differences are $−0.2$~m\AA\ and 7.3~m\AA\, respectively.  Credit: \citeauthor{Borrero2003}, A\&A, 404, 749, 2003, reproduced with permission \textcopyright\ ESO.}
\label{fig:borrero_wavelengths}     
\end{figure}

\begin{figure}
\includegraphics[width=\textwidth]{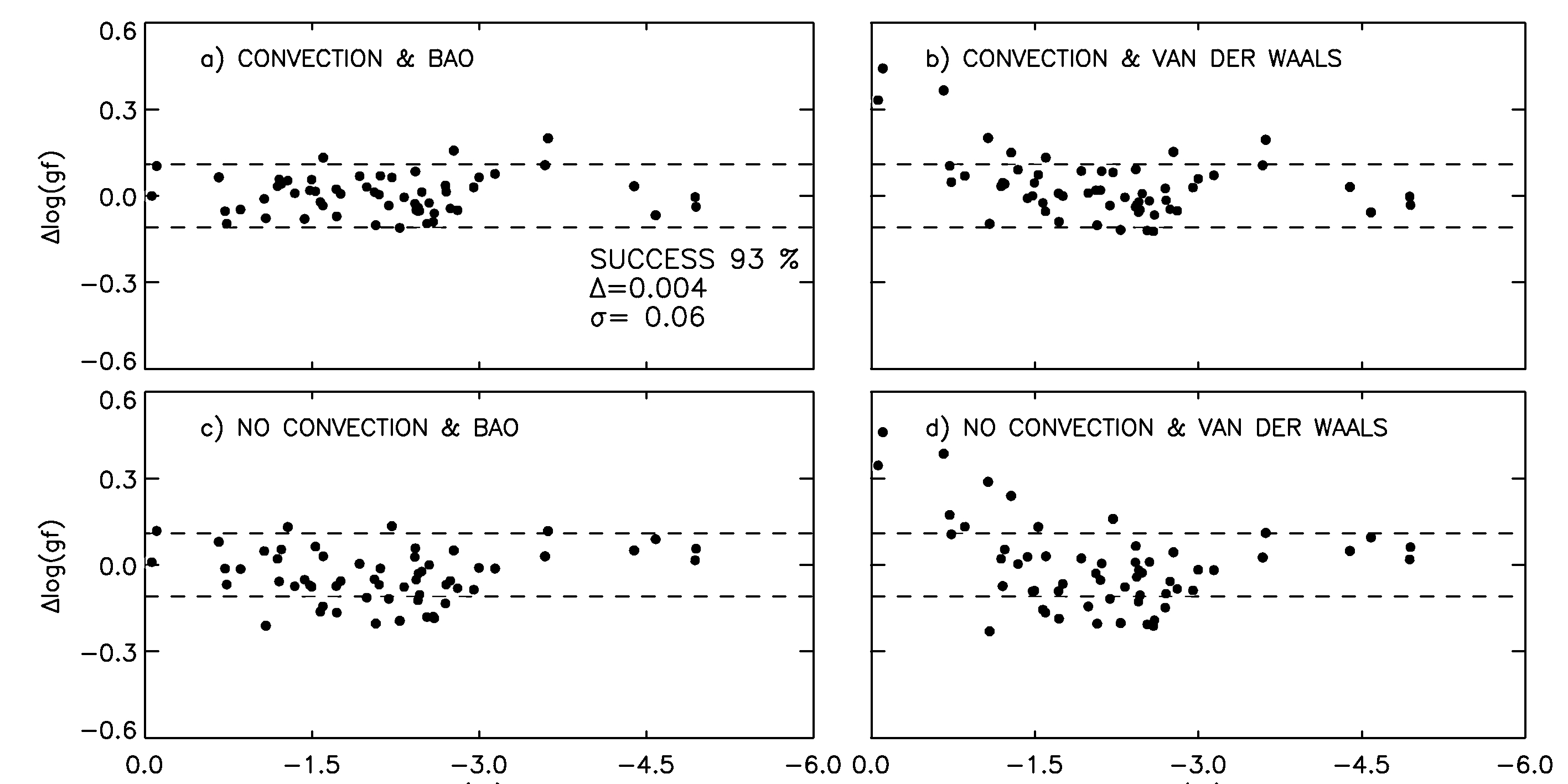}
\caption{a) Comparison between the oscillator strengths of 60 visible \ion{Fe}{i} lines determined from the solar spectrum and from laboratory measurements. The rms difference turns out to be 0.065~dex. The horizontal lines indicate differences of $\pm$0.1~dex. The effects of using a simplified procedure for estimating $gf$-values from the solar spectrum are displayed in the other panels. b) Solar determinations based on classical van der Waals broadening theory(see \S~\ref{sec:metal}, equivalent to LFU theory) instead of data from ABO theory (denoted BAO in figures). c) Solar determinations based on quantum mechanical broadening parameters but no account of convective motions (the model velocities are artificially set to zero). d) Solar oscillator strengths based on van der Waals broadening and no account of convective velocities.  Credit: \citeauthor{Borrero2003}, A\&A, 404, 749, 2003, reproduced with permission \textcopyright\ ESO.}
\label{fig:borrero_loggfs}     
\end{figure}

In addition to identifying levels, if one assumes a good knowledge of spectral line formation in stellar atmospheres (a significant ``if''), then potentially high-quality stellar spectra can be used to determine accurate $gf$-values, in addition to wavelengths and energy levels.  Early examples of large numbers of so-called astrophysical $gf$-values are given by \cite{gurtovenko_establishment_1981-1, gurtovenko_establishment_1982}, who derived $gf$-values for of order of a thousand \ion{Fe}{i} lines, and \cite{thevenin_oscillator_1989,thevenin_oscillator_1990} who derived $gf$-values for some 6000 lines.  Both analyses are based on the solar spectrum and use classical 1D LTE line formation.  Later analysis of these results by \citet[see fig.~2]{Borrero2003} for 80 \ion{Fe}{i} lines in the visible part of the spectrum with laboratory measurements showed that in both cases less than 50\% of the lines were reproduced within the uncertainty of the laboratory measurements.  \cite{Borrero2003} performed an analysis of 83 lines of 6 elements in the near infrared, with the main difference being that a two-component semi-empirical model, and data for collisional broadening due to hydrogen atoms from the ABO theory (to be discussed further in \S~\ref{sec:metal}), are used.  As shown in figs.~\ref{fig:borrero_wavelengths} and~\ref{fig:borrero_loggfs}, using this method, central wavelengths and $gf$-values could be derived with accuracies similar to laboratory measurements.  Fig.~\ref{fig:borrero_loggfs} demonstrates the particular importance of accounting correctly for broadening of spectral lines by collisions with hydrogen atoms in achieving such accuracy.

Generally, isotopic and hyperfine effects can be ignored in analysis of stellar spectra; however, for some elements (e.g. Mn, Co, Cu, Ba), these effects must be accounted for in the analysis of lines that are not in the weak line regime (i.e. have some degree of saturation).  Progress is being steadily made, and we note some examples of measurements of hyperfine \citep{pickering_measurements_1996, blackwell-whitehead_hyperfine_2005, wood_improved_2014} and isotopic \citep{litzen_spectrum_1993, engleman_pd_1998, johansson_experimental_2003} structure relevant to analysis of FGK stars.  For an older review of the role of isotopic and hyperfine splitting in stellar spectra, see \cite{Kurucz1993}.

The importance for astrophysical spectroscopy of critical compilations and databases of fundamental data such as energy levels, wavelengths and transition probabilities, cannot be overstated.  In particular, the NIST Atomic Spectra Database (ASD) \citep{ASD}, the Kurucz database \citep{kurucz_web}, VALD \citep{piskunov_vald_1995,kupka_vald2_1999,ryabchikova_major_2015} and VAMDC \citep{walton_vamdc_2011}, are of great importance to stellar physics in providing large, coherent datasets for modelling stellar atmospheres and their spectra. 

Finally, we note that many analyses are still performed in LTE and thus depend on the availability of partition functions derived from energy levels.  This data, particularly for heavy elements such as the lanthanides and actinides, where data have been rather uncertain, has also seen advances in recent times.  To give one example, in \cite{Christlieb2004a} it was noted that the partition function for thorium from  \cite{morell_age_1992} is around a factor 4 larger than that found in \cite{Irwin1981}, translating to abundance differences of 0.5 dex!  Updated data for neutral, singly- and doubly-ionised species of elements from H to U has recently been calculated by \cite{barklem_partition_2016} using data from the ASD. These calculations are in good agreement with calculations by Lawler and Sneden (personal communication) that have been incorporated into the MOOG code \citep{sneden_nitrogen_1973,sneden_moog_2012}, and have been particularly important for analysis of the heavy $s$- and $r$-process elements.  

\section{Photoionisation cross sections}
\label{sec:photo-ion}

In LTE, bound-free (and free-free) absorption affects abundances only indirectly either through continuous opacity, or via the calculated model atmosphere structure.  However, in non-LTE the photoionisation cross sections are of critical importance.  If in the regions of the atmosphere where photoionisation is important the true radiation field, as given by the mean intensity $J_\nu$, is greater than that expected in LTE, namely that given by the local value of the Planck function $B_\nu$, then the photoionisation rates will be greater than those expected in LTE, leading to so-called ``overionisation''.   This is particularly relevant in FGK stars for neutral species, as they are often the minority species, and generally have ionisation energies corresponding to bound-free edges in the UV, where there is considerable flux in FGK stars.  Thus, accurate photoionisation cross sections for neutral species, and to a lesser degree singly-ionised species, are a requirement for accurate chemical abundances in FGK stars.  We note that cross sections for partial processes from both ground and low-lying excited states are usually a minimum requirement for detailed non-LTE modelling.

The $R$-matrix method \citep{Burke1971,Burke1993} has been of inestimable importance in supplying a unified method for treating atomic processes involving electron-ion-photon interactions, namely electron impact excitation, photoionisation, autoionisation, dielectronic recombination and radiative recombination.  Extensive codes have been developed for $R$-matrix calculations, some of which are freely available \citep[e.g.][]{Berrington1995}.  The $R$-matrix calculations require extensive atomic structure calculations, and these are usually performed using configuration interaction methods in various coupling schemes with various treatments of relativistic effects \citep[e.g.][and subsequent developments]{eissner_techniques_1974,Hibbert1975,dyall_grasp_1989}.  An extensive overview of the various methods and codes is given in the much needed textbook on Atomic Astrophysics and Spectroscopy by \citet[chap 3.]{Pradhan2011}.

The Opacity Project \citep{opacity_project_1995} makes use of these atomic structure calculations and the $R$-matrix method, to calculate a range of data useful for astrophysics, including photoionisation cross sections.  These data are made available via TOPbase \citep{cunto_topbase_1993}, which provides data for all ionisation stages of all elements from H to Si, and elements with even atomic numbers up to Ca (i.e. S, Ar and Ca).  In the time since, additional data have been calculated and made available via the NORAD-Atomic-Data repository \citep{NORAD}, including data for \ion{Ti}{i}, \ion{Cr}{i}, \ion{Fe}{i}.  The data for \ion{Ti}{i} are very recent, from \cite{nahar_photoionization_2015}.  The data for \ion{Cr}{i} are from \cite{nahar_photoionization_2009}, and have been applied to line formation studies by \cite{bergemann_chromium_2010}.  The data for \ion{Fe}{i} is from \cite{bautista_photoionization_1995} and \cite {bautista_atomic_1997}, and it is noted that updated data are available from M.\ Bautista.  Such data has been used in line-formation studies by \cite{Bergemann2012} and \cite{Lind2012}.  Data for \ion{Sr}{i} have been used in \cite{bergemann_nlte_2012}, and are available from the authors.  
 
Photoionisation cross sections, particularly for complex atoms, often show extensive resonance structure due to excitation of core electrons and coupling between photo-excitation of these electrons and that of valence electrons, i.e. due to the existence of autoionising states.  The physics of these resonances structures is explained in detail by \citet[chap 6.]{Pradhan2011}, including a number of illustrative examples.  Of particular interest to the study of non-LTE line formation is the presence of significant resonance structures near threshold.  For example, the results of \citet[][figs.~2 and 3]{bautista_atomic_1997} for \ion{Fe}{i} show such structure.   These resonances are described by Fano profiles \citep{fano_effects_1961, prats_metastable_1964, fano_line_1965}, and usually lead to enhanced absorption corresponding to an autoionising line, although in principle they may also lead to diminished absorption (i.e. a ``hole'' in the background).  The enhanced cross sections can drastically increase the photoionisation rates compared to if only the background is considered, and thus it is important that they are studied and included in the modelling.  However, it should be noted that due to the uncertainties in atomic structure calculations the positions of resonances are often rather uncertain.  \cite{bautista_resonanceaveraged_1998} have suggested a procedure for averaging over resonances in photoionisation cross section calculations in order to minimise errors in the ionisation rates resulting from uncertainties in computed positions of resonances, though still preserving the overall resonant contribution to the cross sections in the important near-threshold regions.  Such resonances are often included in model atmosphere opacity and spectrum synthesis calculations \citep[e.g.][]{kurucz_why_1990, gray_spectrum_1999} as autoionising lines  using the parameterisation of \cite{shore_analysis_1967, shore_scattering_1967}, though often the parameters are unavailable or uncertain.

\section{Collision processes}
\label{sec:collisions}

The effects of the environment, in the form of neighbouring particles, on the atoms absorbing and emitting spectral lines, is of significant importance for analysis of stellar spectra.  Collisions affect the spectral line shapes via collisional broadening.  They also affect the populations of atomic states via inelastic processes.  Both of these processes in turn affect the strength of the observed spectral line, and thus understanding of and the ability to calculate these processes is vital for accurate measurements of stellar properties, especially abundances.

The typical FGK star photosphere consists predominantly of hydrogen atoms, with helium atoms having an abundance an order of magnitude lower.  Atoms with relatively low ionisation potentials are predominantly ionised, yet due to the relative paucity of metal atoms, electrons are rather scarce.  In solar metallicity stars there is roughly only one electron for every $10^4$ hydrogen atoms, and in metal-poor stars this number may grow to as much as $10^6$.  Thus, neutral hydrogen atoms are the dominant perturber \emph{by number}, by a significant margin.  However, as we will, see for certain processes, charged perturbers, namely ions and electrons, may be important, even dominant, due to their stronger interactions and, in the case of electrons, greater speeds. 

\subsection{Broadening of spectral lines}
\label{sec:broadening}

The wings of strong lines of metals (i.e. not hydrogen or helium) are formed due to collisional broadening, and thus are an important diagnostic of pressure (i.e. surface gravity).  They are also very important abundance indicators, since strong lines typically have the most accurate laboratory transition probabilities.  The wings of strong lines also form rather deep in the atmosphere, and therefore due to effective collisional coupling might reasonably be expected to be relatively unaffected by non-LTE effects; however, this should always be checked via detailed modelling.  

The hydrogen Balmer lines present a case of special interest, not least because the $gf$-values are known exactly.  However, the main difference in FGK stars is that since the continuous opacity is due to H$^-$, there is no sensitivity to abundance.  However, due to the 10.2~eV excitation of the lower level, the wings are excellent temperature diagnostics.  The cases of metal and hydrogen lines must also be separated for reasons of atomic physics.  The hydrogen atom energy levels show the ``accidental'' degeneracy with orbital angular momentum quantum number, a consequence of the pure Coulomb potential.  Further, atomic hydrogen is an important perturber itself, resulting in a further degeneracy in hydrogen-hydrogen collisions.  These degeneracies lead to lower order interactions than what is usually found in other atoms with non-degenerate states.  Thus we must examine these two cases separately.

Unfortunately, the general problem of collisional broadening of spectral lines is a time-dependent many-body problem and cannot be solved exactly.  Line broadening theories and calculations often involve a raft of different approximations, and there is a long and complex literature on this subject.  A number of excellent summaries on the subject are available, including the books by \cite{griem_spectral_1974, Griem1997}, the overview articles by \cite{Baranger1962}, \cite{peach_theory_1981, Peach2006}, \cite{lewis_collisional_1980} and \cite{allard_effect_1982}; the text books by \cite{Pradhan2011} and \cite{hubeny_theory_2014} also provide reasonably easy to follow overviews of the main concepts and approximations.

Central to any discussion of collisional broadening are the two limiting approximations: the impact theory and the quasi-static approximation.  In the quantum mechanical impact theory of \cite{Anderson1949, anderson_method_1952} and \cite{baranger_general_1958, baranger_problem_1958, baranger_simplified_1958}, the main assumptions are the \emph{impact approximation} and the \emph{completed collision assumption}.  Note, a distinction is made between the impact \emph{theory} and impact \emph{approximation}; see, e.g., \cite{smith_unified_1969}.  The impact approximation requires that strong collisions do not overlap in time, and a weak collision overlapping a strong one may be neglected by comparison.  Since weak collisions may be treated as separate perturbations, the impact approximation effectively means that all collisions may be treated as isolated events (i.e. a Markovian process), leading to one of the main results of the impact approximation, that the broadening of the line is directly proportional to the perturber density.   This essentially requires that the average collision is weak, which in quantum mechanical terms means that the product of the average interaction energy and a typical collision time must be small compared to $\hbar$, which gives
\begin{equation}
\Delta E \Delta t \ll \hbar \rightarrow \tau \frac{\bar{V}}{\hbar}  \ll 1,
\end{equation}
where $\bar{V}$ is the average absorber-perturber interaction and $\tau$ is a typical collision time.  The impact \emph{theory} extends the impact approximation further by making the completed collision assumption.  This requires that collisions (atom-perturber interactions) occur on short timescales compared to radiative timescales (atom-radiation interactions), such that these interactions are decoupled; in other words, the collision is completed before the absorption of a photon is completed.  Thus, the collision time $\tau$ must be short compared with the timescale characteristic of the atom-photon interaction, which is estimated by $1/\Delta\omega$, where $\Delta\omega$ is the angular frequency of the photon relative to the natural frequency of the transition.  The requirement for the completed collision assumption, and thus the impact theory becomes
\begin{equation}
\tau \, \Delta\omega  \ll 1.
\end{equation}
Under these conditions, the impact theory obtains the well-known Lorentzian line shape, which is valid near line centre, small $\Delta\omega$.   The opposite extreme is the quasi-static approximation in which collisions occur on very long timescales, such that strong collisions are always happening and collisions are never completed during the absorption of a photon.  In this case the motion of perturbers may be disregarded, and the perturbers are assumed to be at fixed positions producing a spectrum of sharp spectral lines.  Statistical averaging over all possible positions produces the broadened spectral line.  This approximation has the opposite validity condition to the impact theory, and thus is always valid in the far line wings.

In FGK star atmospheres, the impact theory is generally valid across the interesting part of the line profile for light, fast moving perturbers such as hydrogen atoms and electrons where typically for metals the absorber-perturber interactions are relatively weak.   In the case of the absorber being hydrogen (or helium), where there is the accidental degeneracy of states leading to stronger interactions with perturbers either via the linear Stark effect for charged particles, or via resonance interactions with hydrogen atoms, the situation is less clear.  For perturbations by protons and ions, the quasi-static approximation is valid across much of the line profile.  For electrons, however, the impact theory is valid near line centre, but breaks down in the wings.  This is due to the breakdown of the completed collision assumption.  Thus, to calculate the broadening of hydrogen lines due to electrons across the entire line profile requires a so-called ``unified line-profile'' approach \citep[see, e.g.][for discussion]{smith_unified_1969,smith_analysis_1973}.  This name originates historically from the fact that early theory could not calculate across the entire line profile from line centre to the far line wings.  Until the advent of the unified theory for electrons, this problem was often circumvented by the use of modified impact theories, such as the introduction of the Lewis cutoff \citep{lewis_stark_1961}.  The unified theory of \cite{smith_unified_1969,Vidal1970, Vidal1971, Smith1972} makes only the impact approximation, but not the completed collision assumption, and corresponds to the impact theory near line centre and to the one-perturber or nearest neighbour quasi-static theories in the far line wings.  In the case of perturbations of absorbing hydrogen atoms by other hydrogen atoms, the resonance interaction dominates, but is not so strong as to totally invalidate the impact theory; however, the impact theory is questionable in the wings of the observable line.

In the following, we will return to these various broadening mechanisms one by one, and will review and discuss the state of the art regarding calculations, and will comment on the applications in modelling FGK stellar spectra, where appropriate.

\subsubsection{Hydrogen Balmer lines}

As mentioned the collisionally broadened wings of Balmer lines are excellent temperature diagnostics for FGK stars, and accurate effective temperatures are a prerequisite for accurate chemical abundances and ages.  Although they were one of the very first spectral diagnostics, and played an important role in early stellar classification, their use as a \emph{precise} diagnostic is a recent development.   This from the difficulty of precisely measuring wide line profiles and correctly placing the continuum.  For this reason stellar effective temperatures have earlier been predominantly determined via colours, sometimes also by making use of hydrogen lines. The Str\"omgren-Crawford photometric system \citep{stromgren_twodimensional_1956, crawford_twodimensional_1958} for example makes use of H$\beta$ filters.  But, by use of well-calibrated CCD-spectra, the detailed analysis of Balmer line wings has recently become a serious competitor to colours as an accurate $T_\mathrm{eff}$ diagnostic, \citep[e.g.][]{fuhrmann_balmer_1994,Barklem2002,Cayrel2011}. In \cite{Barklem2008}, I reviewed various aspects regarding hydrogen lines, including the pros and cons of their use as temperature diagnostics and the status of the understanding of their formation.  This included a discussion of the collisional broadening of these lines, which is revisited and updated here.  

In the broadening of hydrogen lines, the dominant broadening processes are expected to be collisions with charged particles (Stark broadening by electrons and ions) and other hydrogen atoms (self broadening).  As mentioned, in both these processes, the interactions involve degeneracies that complicate the problem compared to the case of lines of metals.  The degeneracy of the energy levels of the hydrogen atom, and thus the linear Stark effect must be considered, and in turn this larger interaction strength leads to the breakdown of the impact theory usually valid for non-degenerate cases.  In the case of self broadening, the relevant degeneracy occurs because the two interacting atoms are indistinguishable, and thus degenerate with respect to a transfer of excitation.  This means that the interaction is stronger than the usual van der Waals type interaction between neutral particles, and gives the so-called resonance interaction in addition to the van der Waals interaction at long range.  This stronger interaction is not so strong as to lead to the total breakdown of the impact approximation, though it does make it more questionable in the far line wings.

The importance of Stark broadening for hydrogen lines has been known for almost a century.  Already in \cite{struve_stark_1929} the noticeably larger widths of hydrogen lines in stellar spectra were explained as due to the linear Stark effect, and since that time, our understanding of Stark broadening of hydrogen lines has advanced considerably.  Prior to \cite{griem_starkeffektverbreiterung_1954}, it was generally thought that electrons were unimportant, and that the dominant contribution was due to interactions with ions.  Quantitative comparison of experiment and theory showed this to be incorrect, the electron contribution being comparable.  Later work by Griem and collaborators \citep[e.g.][]{Griem1959,kepple_improved_1968} produced theories treating the quasi-static ions and impact electrons.  However, as discussed above, the impact theory is not valid for electron collisions in the far wings of hydrogen lines.  Unified theory calculations were produced by \cite{Vidal1973}, treating the ions in the quasi-static approximation.  These calculations have been of fundamental importance in the modelling of stellar atmospheres and in the accurate interpretation of stellar spectra.  Later improvements were brought about by the introduction of the model microfield method \citep{frisch_theory_1971, brissaud_theory_1971}, which further relaxes the quasi-static approximation and thus includes ion dynamics.  Calculations using this method have been done by \cite{Stehle1994}, \cite{Stehle1999}, and \cite{Stehle2010} and compare very well with experiments of \cite{ehrich_experimental_1980} \citep[see][]{Stehle1994}.  The differences compared to the calculations by \cite{Vidal1973} are small and generally occur in the line cores. Thus, while important for comparison with laboratory data \citep{Cooper1974}, these differences are lost in astrophysical spectra as this region of the line profile is dominated by Doppler broadening, and the cores are not used in effective temperature determinations \citep[see, e.g.][]{Lemke1997}.

The importance of self broadening due to the perturbation of absorbing hydrogen atoms by other hydrogen atoms has only been fully appreciated relatively recently, particularly in the context of metal-poor stars.  In addition to the usual van der Waals type interactions between neutral particles, with long range attractive $R^{-6}$ behaviour, a hydrogen atom in a $p$-state undergoes also a resonance interaction with hydrogen atoms in their ground state, leading to two possible states with an attractive or repulsive interaction with a stronger $R^{-3}$ behaviour.  This resonance interaction may be understood quantum mechanically in terms of that a hydrogen atom in an $np$-state may emit a Lyman series photon that can be directly absorbed by the ground state hydrogen atom.  This permits an interaction in first order, whereas by comparison the van der Waals interaction requires the involvement of two photons and is a second order process.  This interaction may also be understood classically, in terms of two oscillating dipoles with the same natural frequencies \cite[][p.\ 476-478]{bohm_quantum_1951}.  \cite{cayrel_zur_1960} pointed out the importance of resonance broadening for H$\alpha$ in the sun.  Note that here the nomenclature ``resonance broadening'' refers specifically to broadening due to the resonance interaction.  ``Self broadening'' refers to broadening due to all types of interactions between two hydrogen atoms, including the resonance interaction, but also the dispersive and inductive interactions making up the van der Waals interactions.  Theoretical treatments of resonance broadening in the impact approximation based on a $C/R^3$ form of the potential are in very good agreement.  The most widely used is perhaps \cite{ali_theory_1965, ali_theory_1966}.  It was later pointed out by \cite{lortet_broadening_1969} that van der Waals type interactions will be important, though these authors realised that the theory they used for calculating the van der Waals interaction based on a multipole expansion of the interaction was inadequate.  \cite{Barklem2000,Barklem2000a} developed a theory of self broadening including van der Waals type interactions, without resort to a multipole expansion but instead using perturbation theory to compute the interaction potentials, and calculations were performed for the three lowest Balmer lines.  The impact of these calculations was shown to be important, particularly in metal-poor stars, resulting in cooler effective temperatures by of order 100~K.  This theory was developed in the impact theory, but the authors noted that this approximation was likely to break down in the far wings.  Later, \cite{Allard2008} performed calculations using the unified theory, based on quantum chemical data for the H$_2$ molecule by \cite{spielfiedel_initio_2003}.  The calculations included the effect of variation of the transition moment, an improvement made possible by earlier theoretical work \citep{allard_effect_1999}.  Due to lack of input data, these calculations have at present only been done for H$\alpha$.  They show quite reasonable agreement with the earlier \cite{Barklem2000a} calculations, to within around 5\% in the line widths, and further demonstrate that the impact theory is probably reasonable out to around 20~{\AA} from line centre, thus covering most of line profile that is useful as a stellar diagnostic.  It would be important to extend the unified theory calculations with quantum chemistry type potentials and transition moments to the higher Balmer lines.   

At this point it is worth noting that the accurate modelling of the broadening of hydrogen lines in stellar atmospheres, including the Balmer lines, requires a very significant amount of information regarding their physics, including the various broadening mechanism.  In order to make this easier, \cite{barklem_hydrogen_2003, barklem_hlinop_2015} have made a freely available package of codes for the calculation of hydrogen line opacities.  The package \texttt{HLINOP}, is a collection of codes for computing hydrogen line profiles and opacities in the conditions typical of stellar atmospheres.  The package includes a main code, also called \texttt{HLINOP}, for calculating any line of neutral hydrogen (suitable for model atmosphere calculations), based on the Fortran code by Kurucz and Peterson found in the \texttt{ATLAS} model atmosphere programs \citep{kurucz_atlas12_2013}. For Stark broadening, rather than using detailed tabulations of line profiles which is computationally expensive, it uses wing-formulae from \cite{Griem1959} and \cite{griem_wing_1962}, adjusted to better match the \cite{Vidal1973} calculations.  The package also includes \texttt{HLINPROF}, for detailed, accurate calculation of lower Balmer line profiles (suitable for detailed analysis of Balmer lines).  Also included is a wrapper code \texttt{HBOP} to implement the occupation probability formalism of \cite{Dappen1987} and thus account for the merging of bound-bound and bound-free opacity.   This can be particularly important in modelling higher series members.

\cite{Barklem2008} also raised the issue of whether, or to what accuracy, the separate treatment of Stark and self broadening is valid.  The Coulomb forces due to charged particles affect the structure of the emitting hydrogen atom and this effect is neglected in treating the broadening mechanisms separately.  In section 5.3 of \cite{Barklem2000a}, this issue was briefly touched upon in the context of use of the ``p-d approximation'', where the electron is considered to be in a central field and thus the wavefunction can be separated into radial and angular parts in spherical polar coordinates.  However, in the presence of an electric field, such spherical symmetry is broken, and parabolic coordinates must be employed.  It was argued there that the quasi-static ion field in cool stars is quite weak and thus the use of the spherical polar coordinates and associated quantum numbers $nlm$ was reasonable.  Calculations to quantify the effect of this would be very worthwhile.  Another issue is the adiabatic approximation, where transitions between different electronic states (corresponding asymptotically to the same principal quantum number of the perturbed atom) are neglected.  In recent work \cite{Barklem2000a} and \cite{Allard2008} have presented arguments why this approximation should be valid.  However, as for the influence of the ion field, quantitative evaluation of the importance of this effect would be important.  Such effects may be estimated to be ``small''; however, our aim is to interpret hydrogen lines in FGK stellar spectra at an accuracy well below the per cent level.

Unfortunately, experimental results for line broadening are rather scarce.  In the case of Stark broadening there are some such as \cite{ehrich_experimental_1980}, but for electron temperatures that are much higher than those relevant for FGK stellar atmospheres.  To our knowledge there are no experiments for self broadening of hydrogen.  This means that one is rather dependent on astrophysical tests to judge if the broadening theories are complete and accurate.  However, there is always the caveat that such tests depend on the accuracy of stellar atmosphere modelling, and any discrepancies may well also lie with the usual suspects: 3D and non-LTE, not to mention other effects such as magnetic fields.  Some recent work has indicated that there may be problems with 1D LTE modelling of Balmer lines, on the basis that temperatures are not in perfect agreement with those from other methods.  Work by  \cite{Barklem2002} found that while for a sample of stars, results agreed reasonably well with those from the Infrared Flux Method results of \cite{alonso_determination_1996}, for the best test case, the sun, the modelling employing late 1990's generation MARCS 1D model atmospheres \citep{asplund_lineblanketed_1997} and LTE line formation, gave an effective temperature 50~K too cool compared with the precisely known true value.  However, the results agreed within the estimated uncertainties.  Later work by \cite{Cayrel2011} compared accurate effective temperatures for a dozen stars found with the direct method of combining the apparent diameter and the bolometric flux, with those derived from H$\alpha$ in 1D LTE modelling.  They found a systematic offset, with H$\alpha$ giving temperatures of order 100~K too cool.  \cite{Norris2013} used spectrophotometric fits and 1D LTE modelling of Balmer lines to derive stellar parameters for a sample of very metal-poor stars.  They also found that Balmer-line temperatures yield systematically cooler values of order 100~K.  Work on modelling Balmer lines in FGK stars in non-LTE \citep{Barklem2007, Mashonkina2008}, 3D \citep{Ludwig2009}, and 3D non-LTE \citep{Pereira2013} is in its infancy but initial results seem promising with corrections generally going in the direction of improving agreement for standard stars with well determined temperatures from other methods, particularly for the sun.

Finally, it should also be noted that there has been much advance in recent times in calculating the effects on the far wings of Lyman $\alpha$ due to collisions by protons and hydrogen atoms, e.g. \citep{allard_new_1998, Allard2009}.   The unified theory, enabling the line profile to be calculated from line centre to the far wing, including accounting for the variation of the transition moment, is of particular importance here \citep{allard_effect_1999}.  It should also be noted that laboratory measurements in laser-induced plasma play an important role here \citep{kielkopf_observation_1998, Allard2009}.   This work is mostly of importance in modelling stellar atmospheres, as the red wing of Lyman-$\alpha$ contributes substantial opacity, which indirectly may affect derived abundances and properties of FGK stars.  These calculations have been implemented in the \texttt{ATLAS} codes \citep{castelli_ultraviolet_2001}, and this implementation is also included in \texttt{HLINOP}.

\subsubsection{Metal lines}
\label{sec:metal}
 
For lines of metal atoms and ions perturbed by electrons and hydrogen atoms the interaction strengths are sufficiently weak and the velocities in stellar atmospheres are sufficiently high, so that the impact theory remains valid across the entire line profile, even for the very strongest and broadest lines.  Electrons interact with atoms via the quadratic Stark effect with a long-range dependence on distance of $R^{-4}$; in the case of ions there are stronger Coulomb terms, but as these terms have no state dependence they do not affect broadening, except via dynamical effects.  Hydrogen atoms interact with other atoms at long range via the attractive van der Waals force with a dependence on distance of $R^{-6}$; in the case of ions there are stronger inductive interactions, but again these do not depend on the state of the ion, and thus may only affect broadening via dynamical effects.  In reality, of course the interactions at shorter range depart from these long range behaviours, and as we will see, this usually needs to be accounted for in accurate calculations.

The standard, classical theory for collisional broadening of spectral lines, which is found in almost every textbook on stellar atmospheres \citep[e.g.][]{aller_astrophysics_1963,hubeny_theory_2014} or formulae compilation \citep[e.g.][]{lang_astrophysical_1999} is the Lindholm-Foley theory \citep{lindholm_pressure_1946,foley_pressure_1946}.  This theory makes a number of assumptions, in particular a $C_n/R^n$ form for the interaction potential, straight-line classical trajectories, averaging over magnetic substates, among others, which permits analytical expressions for the cross section and line widths to be obtained.  The Lindholm-Foley theory also predicts a pressure induced line shift, which is seen in experiments.  The results for the various forms of the interaction with different $n$ are given in numerous textbooks, such as those mentioned above.   The key difficulty is the calculation of the constant $C_n$, assuming it is unavailable by other means such as measurements, which is usually the case.  Calculation usually requires a summation over $f$-values and energies for all perturbing states of the atom or ion, and large scale calculations of such data have been undertaken by Kurucz \citep[see, e.g., p.\ 75-76][]{kurucz_semiempirical_1981}.  In the case of the van der Waals interaction, $n=6$, the summation can be approximated by noting that terms related to the hydrogen atom dominate the sums, giving a simple expression for $C_6$ in terms of the static dipole polarisability of hydrogen in its ground state, which is known precisely, and the mean-square position of the valence electron in the atom or ion, which can be easily estimated from hydrogenic formulae \citep{unsold_physik_1955}.  This is usually referred to as the Uns\"old approximation, and together with the Lindholm-Foley theory leads to simple formulae for the collisionally broadened line width.  This theory will be referred to as Lindholm-Foley-Uns\"old (LFU) theory.  

Before discussing the collisional broadening due to hydrogen atoms further, some comments on Stark broadening are warranted.  Calculations almost universally show that in FGK stars, collisional broadening by hydrogen atoms is more important than quadratic Stark broadening.  For this reason our discussion in this section will focus on broadening by hydrogen collisions.  However, for maximum accuracy Stark broadening should always be accounted for where possible.  This is relatively straightforward due to the efforts of Sahal-Br\'echot and collaborators in performing large-scale calculations, which are made readily available via the STARK-B database \citep{sahal-brechot_starkb_2015}.  These calculations are based on theoretical considerations reviewed in \cite{sahal-brechot_widths_2014}, which are developments of earlier work \citep{sahal-brechot_impact_1969,sahal-brechot_impact_1969-1}.  It should be further noted that there can be exceptions where Stark broadening may be very important even in relation to broadening by hydrogen atom collisions.  One example was found by \cite{Anstee1997}, where transitions of \ion{Fe}{i} involving the $e^5D$ state were seen to give abundances in disagreement with other lines.  This is explained by the fact that this level has a nearby perturbing level, leading to exceptionally large Stark broadening.

As noted above, the collisional broadening due to hydrogen atoms is generally of the greatest importance in FGK stellar spectra.  The LFU theory is still widely used in astrophysics, though almost always with a correction factor, often derived from spectra of well studied stars, especially the sun.  This is necessary as the LFU theory usually significantly underestimates the broadening.  \cite{holweger_damping_1971} noted the need to arbitrarily increase the $C_6$ values to match the solar spectrum of the \ion{Na}{i} D lines, though it was pointed out by \cite{roueff_broadening_1971} that this was not physical.  This pointed to a deeper problem, namely that the use of a $C_6/R^6$ form of the potential was inadequate.  In fact, \cite{omara_use_1978} showed that multipole expansion of the interaction potential in general is not useful for broadening by hydrogen collisions.  More detailed calculations \citep[e.g.][]{lewis_broadening_1971,anstee_investigation_1991} have shown definitively that interactions at shorter range are important.  At intermediate to short range the electronic wavefunctions of the two atoms start to overlap and thus the van der Waals $C_6/R_6$ interaction, or any multipole interaction for that matter, is not valid and underestimates the interaction.  It is for this reason that the commonly used term ``van der Waals broadening'', is in fact a misnomer, and it is more appropriate to use the description ``collisional broadening by hydrogen atoms''.

Thus, the situation for calculations of various lines of great astrophysical interest has improved significantly.  Tables~\ref{tab:broad_na} and~\ref{tab:broad_ca} collate various data on strong lines of Na and Ca, respectively, from theory, experiment and inferred from the solar spectrum.  The first thing to note is the almost complete absence of experimental data for broadening of spectral lines by atomic hydrogen.  There are two shock-tube experiments for the \ion{Na}{i} D lines, that by \cite{baird_width_1979} and that by \cite{lemaire_broadening_1985}.  These two results disagree significantly, the latter result being almost 60\% greater than the former; however the latter results are claimed to be far more precise than the earlier results.  Thus, the result by \cite{lemaire_broadening_1985} is perhaps the only reliable one.  It has an estimated uncertainty of 14\%, giving a line width per perturber of $w/N=(13.8\pm1.9) \times 10^{-9}$ rad s$^{-1}$ cm$^3$.  This is somewhat higher than most modern calculations, which are in the range 10--$11.4 \times 10^{-9}$ rad s$^{-1}$ cm$^3$.  Line widths inferred from the solar spectrum by \cite{smith_collisional_1985} are in much better agreement with the theoretical values.  This might indicate that the experimental values of \citeauthor{lemaire_broadening_1985} are overestimated; however, given the uncertainties involved the evidence does not rule out a value at the lower extreme of the experimental uncertainty range, around $12 \times 10^{-9}$ rad s$^{-1}$ cm$^3$.  Given the importance of collisional broadening by hydrogen in astrophysics, the experimental situation is unsatisfactory, and further experiments with modern methods would be most welcome.  The lack of experiments can be partially explained by the difficulties in working with atomic hydrogen at the temperatures of interest; however, the problems are clearly not insurmountable, as evidenced by the two existing experiments.  It should be mentioned that there are many more experimental results for collisional broadening by noble gases, as they are far more easy to work with, and there have been various attempts to extrapolate these results to broadening by hydrogen \citep[e.g.][]{oneill_collisional_1980}.  Such attempts have often made such extrapolations on the basis of the LFU theory using long-range potentials, which has been seen above to be inadequate.  Given that interatomic potentials at shorter internuclear distances depend non-trivially on the involved perturber, there is no obvious way to accurately relate results for rare gases to the case of hydrogen.

\begin{table}
\caption{Line widths $w$ and shifts $d$ due to collisions with neutral hydrogen at 5000~K for \ion{Na}{i} lines of interest in FGK stars.  The widths and shifts are presented per unit perturber density.  The parameters $\beta$ are the temperature coefficients, assuming $w \propto T^{\beta_w}$ and $d \propto T^{\beta_d}$.}
\label{tab:broad_na}       
\begin{threeparttable}
\scriptsize
\begin{tabular}{lrr|rr|rr|rr}
\hline\noalign{\smallskip}
Source & \multicolumn{2}{c|}{$w/N$} & \multicolumn{2}{c|}{$\beta_w$} &  \multicolumn{2}{c|}{$d/N$} & \multicolumn{2}{c}{$\beta_d$} \\
       & \multicolumn{2}{c|}{[$10^{-9}$ rad s$^{-1}$ cm$^3$]} & & & \multicolumn{2}{c|}{[$10^{-9}$ rad s$^{-1}$ cm$^3$]} &  &  \\
\noalign{\smallskip}\hline\noalign{\smallskip}
&&&&&&&&\\
\multicolumn{5}{c}{\underline{\ion{Na}{i} D lines: $3s$--$3p$ }} \\
&&&&&&&&\\
       & D$_1$ & D$_2$ & D$_1$ & D$_2$& D$_1$ & D$_2$& D$_1$ & D$_2$     \\
\emph{Experiment}: & & & & & & & & \\
&&&&&&&&\\
\cite{baird_width_1979}        &  \multicolumn{2}{c|}{8.8}             & \multicolumn{2}{c|}{$-0.3\pm0.3$} &                                &    &    &   \\
\cite{lemaire_broadening_1985} &  \multicolumn{2}{c|}{$13.8\pm1.9$}   & \multicolumn{2}{c|}{$0.42$}       & \multicolumn{2}{c|}{$-0.52\pm0.16$} &    &   \\
&&&&&&&&\\
\emph{Theory}: & & & & & & & & \\
&&&&&&&&\\
\cite{unsold_physik_1955} (LFU)     &   \multicolumn{2}{c|}{6.0}             & \multicolumn{2}{c|}{$0.30$}        & \multicolumn{2}{c|}{$-2.16$}         & \multicolumn{2}{c}{$0.30$}  \\ 
\cite{lewis_broadening_1971}  &   \multicolumn{2}{c|}{8.3}             & \multicolumn{2}{c|}{$0.42$}       & \multicolumn{2}{c|}{$-0.60$}         &     &   \\ 
\cite{Roueff1974}             &     $8.57$        &     $8.35$         & $0.38$     &    $0.44$            & $-0.67$       &   $-0.61$            &    0.48 & 0.99       \\ 
\cite{vanrensbergen_broadening_1975}&   \multicolumn{2}{c|}{7.9}       & \multicolumn{2}{c|}{$ $}          & \multicolumn{2}{c|}{$-0.15$}         &     &   \\ 
\cite{OMara1976}\tnote{a}     &   \multicolumn{2}{c|}{14.4}            & \multicolumn{2}{c|}{$0.37$}       & \multicolumn{2}{c|}{$-5.89$}         &  \multicolumn{2}{c}{$0.40$}     \\ 
\cite{monteiro_broadening_1985}\tnote{b} &   \multicolumn{2}{c|}{$10.6^{+0.2}_{-0.3}$} & \multicolumn{2}{c|}{$0.41$} & \multicolumn{2}{c|}{$-1.08$}         &  \multicolumn{2}{c}{$ $}     \\\cite{Krsljanin1993}        &   \multicolumn{2}{c|}{$9.72$} & \multicolumn{2}{c|}{$0.36$} & \multicolumn{2}{c|}{$-1.12$}         &  \multicolumn{2}{c}{$0.33$}     \\     
\cite{Anstee1995}\tnote{c}    &   \multicolumn{2}{c|}{$11.7^{+0.0}_{-0.7}$} & \multicolumn{2}{c|}{$0.38$}  & \multicolumn{2}{c|}{$-2.82^{+0.6}_{-0.0}$}         &  \multicolumn{2}{c}{$0.33$}     \\ 
\cite{Leininger2000}\tnote{d} &   \multicolumn{2}{c|}{$11.0$}          & \multicolumn{2}{c|}{$0.40$}       & $-0.70$ & $-0.55$         &  \multicolumn{2}{c}{$ $}     \\ 
\cite{Kerkeni2004}            &   11.4   &    11.3                     & \multicolumn{2}{c|}{$0.36$}  &&                                    &&  \\
\cite{peach_recent_2011}      &   \multicolumn{2}{c|}{10.57}           & \multicolumn{2}{c|}{$0.40$}       & \multicolumn{2}{c|}{$-1.165$}         & \multicolumn{2}{c}{$0.16$}  \\ 
&&&&&&&&\\
\emph{Solar}: & & & & & & & & \\
&&&&&&&&\\
\cite{smith_collisional_1985} &   $11.7\pm2.1$   &   $10.6\pm2.1$       & \multicolumn{2}{c|}{$ $}       & \multicolumn{2}{c|}{$$}         & \multicolumn{2}{c}{$$}  \\
&&&&&&&&\\
\multicolumn{5}{c}{\underline{\ion{Na}{i} near IR triplet : $3p$--$3d$ }} \\
&&&&&&&&\\
\cite{unsold_physik_1955} (LFU)     &    \multicolumn{2}{c|}{$10.6$}        & \multicolumn{2}{c|}{$0.30$}       & \multicolumn{2}{c|}{$$}         & \multicolumn{2}{c}{$$}  \\
\cite{Barklem1997a}           &    \multicolumn{2}{c|}{$22.8$}        & \multicolumn{2}{c|}{$0.37$}       & \multicolumn{2}{c|}{$$}         & \multicolumn{2}{c}{$$}  \\
\cite{Leininger2000}\tnote{e}          &    \multicolumn{2}{c|}{$28.4$}        & \multicolumn{2}{c|}{$0.39$}       & \multicolumn{2}{c|}{$$}         & \multicolumn{2}{c}{$$}  \\
&&&&&&&&\\
\multicolumn{5}{c}{\underline{\ion{Na}{i} visual triplet : $3p$--$4d$ }} \\
&&&&&&&&\\
\cite{unsold_physik_1955} (LFU)     &    \multicolumn{2}{c|}{$20.7$}        & \multicolumn{2}{c|}{$0.30$}        & \multicolumn{2}{c|}{$$}         & \multicolumn{2}{c}{$$}  \\
\cite{Barklem1997a}           &    \multicolumn{2}{c|}{$55.2$}        & \multicolumn{2}{c|}{$0.34$}       & \multicolumn{2}{c|}{$$}         & \multicolumn{2}{c}{$$}  \\
\cite{Leininger2000}          &    \multicolumn{2}{c|}{$18.6$}        & \multicolumn{2}{c|}{$0.14$}       & \multicolumn{2}{c|}{$$}         & \multicolumn{2}{c}{$$}  \\
\noalign{\smallskip}\hline
\end{tabular}
\begin{tablenotes}
\item [a] Calculated according to this theory (Brueckner-O'Mara), but data not presented in paper; \cite[see][]{anstee_investigation_1991}.
\item [b] Range of variation of $w/N$ given for three different sets of potentials.  Behaviour of shift not uniform with temperature.
\item [c] Range of variation of $w/N$ given for different calculations in \cite{anstee_investigation_1991,Anstee1992}.  
\item [d] $D_2$ width slightly larger than $D_1$ .  Shifts read from plot, and behaviour of shift not uniform with temperature.
\item [e] \cite{Sanchez-FortunStoker2003} have also investigated line mixing in these lines, and found the effects to be small.
\end{tablenotes}
\end{threeparttable}
\end{table}

\begin{table}
\caption{Line widths $w$ and shifts $d$ due to collisions with neutral hydrogen at 5000~K for \ion{Ca}{i} and \ion{Ca}{ii} lines of interest in FGK stars.  The widths and shifts are presented per unit perturber density.   The parameters $\beta$ are the temperature coefficients, assuming $w \propto T^{\beta_w}$ and $d \propto T^{\beta_d}$.}
\label{tab:broad_ca}       
\begin{threeparttable}
\scriptsize
\setlength{\tabcolsep}{4pt}
\begin{tabular}{lrrrrrr|r|rrr|r}
\hline\noalign{\smallskip}
Source & \multicolumn{6}{c|}{$w/N$} & $\beta_w$ &  \multicolumn{3}{c|}{$d/N$} & $\beta_d$ \\
       & \multicolumn{6}{c|}{[$10^{-9}$ rad s$^{-1}$ cm$^3$]} &  & \multicolumn{3}{c|}{[$10^{-9}$ rad s$^{-1}$ cm$^3$]} &    \\
\noalign{\smallskip}\hline\noalign{\smallskip}
&&&&&&&&&&&\\
\multicolumn{8}{c}{\underline{\ion{Ca}{i} lines: $4p$--$5s$ }} \\
&&&&&&&&&&&\\
       & \multicolumn{2}{c}{6102 \AA}  & \multicolumn{2}{c}{6122 \AA} & \multicolumn{2}{c|}{6162 \AA} &  &  6102 \AA  & 6122 \AA & 6162 \AA  &    \\
\emph{Theory}: & & & & & & & & & & \\
&&&&&&&&&&&\\
\cite{unsold_physik_1955}  (LFU)            &  \multicolumn{2}{c}{10.4} & \multicolumn{2}{c}{10.4} & \multicolumn{2}{c|}{10.4}  & 0.30 & $-3.7$ & $-3.7$ & $-3.7$ & 0.30 \\
\cite{spielfiedel_collision_1991}     &  \multicolumn{2}{c}{31.0} & \multicolumn{2}{c}{23.6} & \multicolumn{2}{c|}{25.8}  &     & $+3.5$ & $+1.7$ & $+2.2$ &     \\
\cite{Anstee1995}                     &  \multicolumn{2}{c}{25.0} & \multicolumn{2}{c}{25.0} & \multicolumn{2}{c|}{25.0}  & 0.37& $-10.1$ & $-10.1$ & $-10.1$ & 0.43 \\
&&&&&&&&&&&\\
\emph{Solar}: & & & & & & & & & & \\
&&&&&&&&&&&\\
\cite{oneill_collisional_1980}        &   &    &  &   & \multicolumn{2}{c|}{$24\pm6$} & & & & & \\
\cite{smith_nonresonance_1986}        &   &    &  &   & \multicolumn{2}{c|}{$30\pm2$} & & & & & \\
&&&&&&&&&&&\\
\multicolumn{8}{c}{\underline{\ion{Ca}{ii} UV doublet : $4s$--$4p$ }} \\
&&&&&&&&&&&\\
      &  \multicolumn{3}{c}{3933 \AA\ (K)} &  \multicolumn{3}{c|}{3969 \AA\ (H)}  & & & & & \\
\emph{Theory}: & & & & & & & & & & \\
&&&&&&&&&&&\\
\cite{unsold_physik_1955}  (LFU)            & \multicolumn{3}{c}{3.3}  & \multicolumn{3}{c|}{3.3} & 0.30 &&&&\\
\cite{Deridder1976} \tnote{a}         & \multicolumn{3}{c}{5.5}  & \multicolumn{3}{c|}{5.5} &     &&&&\\
\cite{monteiro_broadening_1988} \tnote{b}      & \multicolumn{3}{c}{8.3}  & \multicolumn{3}{c|}{8.2} &     &&&&\\
\cite{Barklem1998a}                   & \multicolumn{3}{c}{6.6}  & \multicolumn{3}{c|}{6.6} &  0.39   &&&&\\
&&&&&&&&&&&\\
\emph{Solar}: & & & & & & & & & & \\
&&&&&&&&&&&\\
\cite{ayres_reexamination_1977}      & \multicolumn{6}{c|}{$8.5\pm2.5$}  &    &&&&\\
&&&&&&&&&&&\\
\multicolumn{8}{c}{\underline{\ion{Ca}{ii} IR triplet : $3d$--$4p$ }}\\
&&&&&&&&&&&\\
       & \multicolumn{2}{c}{8498 \AA}  & \multicolumn{2}{c}{8542 \AA} & \multicolumn{2}{c|}{8662 \AA} &  &   &  &  &    \\
\emph{Theory}: & & & & & & & & & & \\
&&&&&&&&&&&\\
\cite{unsold_physik_1955}  (LFU)            &  \multicolumn{2}{c}{5.0} & \multicolumn{2}{c}{5.0} & \multicolumn{2}{c|}{5.0}  & 0.30 & & & & \\
\cite{Barklem1998a}                   &  \multicolumn{2}{c}{8.2} & \multicolumn{2}{c}{8.2} & \multicolumn{2}{c|}{8.2}  & 0.36 & & & & \\
&&&&&&&&&&&\\
\emph{Solar}: & & & & & & & & & & \\
&&&&&&&&&&&\\
\cite{smith_collisional_1988}         &  \multicolumn{2}{c}{$10.7\pm1.0$} & \multicolumn{2}{c}{$10.3\pm0.5$} & \multicolumn{2}{c|}{$10.0\pm0.5$}  & & & & & \\
\cite{Barklem1998a}                   &  \multicolumn{2}{c}{$8.7\pm0.8$} & \multicolumn{2}{c}{$8.1\pm0.8$} & \multicolumn{2}{c|}{$8.7\pm0.8$}  & & & & & \\
\noalign{\smallskip}\hline
\end{tabular}
\begin{tablenotes}
\item [a] As calculated from their tables by \cite{monteiro_broadening_1988}.
\item [a] Temperature dependence varying.
\end{tablenotes}
\end{threeparttable}
\end{table}

\begin{figure}
\center
\includegraphics[width=\textwidth]{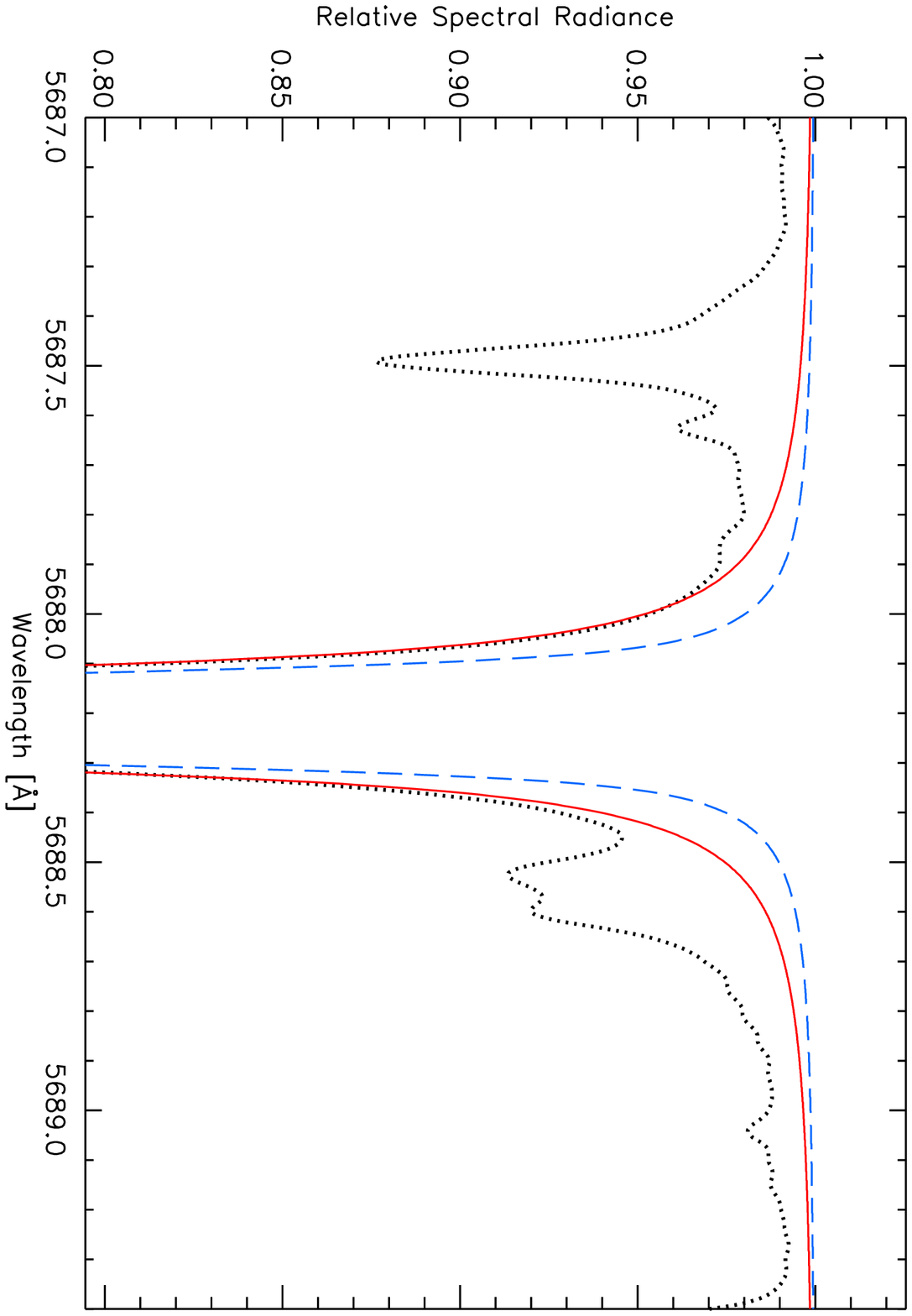}
\caption{Line profiles for the $3p$--$4d$ multiplet at 5688~\AA of \ion{Na}{i}. The observed relative solar radiance (often called ``intensity'' in astrophysics) at disk centre \citep{delbouille_atlas_1990} is the (black) dotted curve. The (red) full curve is the model spectrum using collisional broadening data from \cite{Barklem1997a} and the (blue) dashed curve represents the data from \cite{Leininger2000}. All model spectra assume the abundance of $\log(N_\mathrm{Na}/N_\mathrm{H})+12 =6.27$.  Adapted from \cite{Barklem2001}.}
\label{fig:Na5688}     
\end{figure}

Regarding the other results for Na and Ca, modern theoretical calculations for the line widths are often in reasonable agreement with each other, and with results derived from the solar spectrum, at around the 20-30\% level, and this may give some idea of the uncertainties involved.  One notable exception is the \ion{Na}{i} visual triplet, $3p$-$4d$ where the results from \cite{Barklem1997a} are 2.7 times larger than those from \cite{Leininger2000}.  This disagreement has been discussed by \cite{Barklem2001}, including a comparison of spectral synthesis with the solar spectrum, where the abundance of Na has been assumed to be $\log(N_\mathrm{Na}/N_\mathrm{H})+12 =6.27$, which is in agreement with modern values both for the solar photosphere and for meteorites \citep[e.g.][]{Asplund2009}.  The synthesis and comparison of the 5688~{\AA} line is reproduced in fig.~\ref{fig:Na5688}, and clearly favours the larger value. \cite{Barklem2001} have shown that this may be due to how the effects of the ionic configuration via avoided ionic crossings in the NaH potentials (see \S~\ref{sect:hinelast} and fig.~\ref{fig:crossing}) are included in the calculations, arguing that they should not contribute adiabatically in this case.  The \cite{Barklem1997a} calculations fare better since effects of the ionic configuration are ignored, and \cite{Barklem2001} go on to derive indicative results of when avoided ionic crossings should be important.  Calculations, in particular for these lines, with inclusion of avoided crossings explicitly in the collision dynamics would be very important.  

\begin{figure}
\center
\includegraphics[width=.52\textwidth,angle=90]{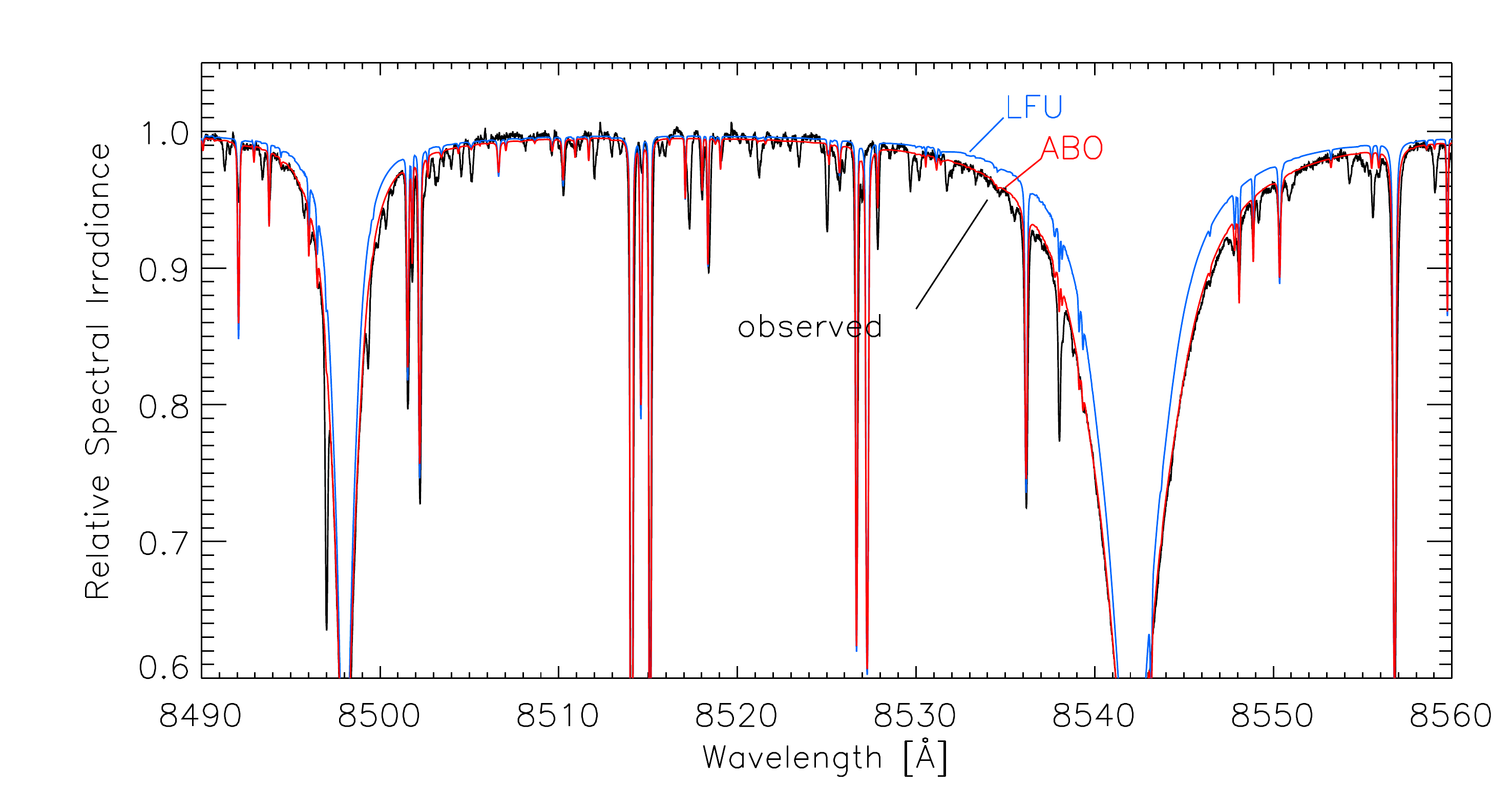}
\caption{Comparison of observed relative solar irradiance (often called ``flux'' in astrophysics) (black) with model spectra using broadening data from LFU (blue) and ABO (red) theories for the \ion{Ca}{ii} infrared triplet.  The adopted abundance is $\log(N_\mathrm{Ca}/N_\mathrm{H})+12 =6.35$, which is the meteoritic abundance from \cite{grevesse_standard_1996}, which is in good agreement with currently accepted solar values.}
\label{fig:abo}     
\end{figure}

Naturally, for very accurate abundances it would be desirable to reduce these uncertainties in the widths, and further experiments to guide theory would be very important in this regard.  The above indicates that the inclusion of effects due to ionic configurations may be an important, as yet unexplored aspect in theoretical calculations, and experiments on this line would be particularly welcome.  For the time being, modelling of FGK stellar spectra is dependent on theoretical calculations.  Tables~\ref{tab:broad_na} and~\ref{tab:broad_ca} cover the lines for which there are detailed, modern calculations.  But to model the very large number of spectral lines in stellar spectra, at present at least, other methods are needed.  This is to a large degree why the LFU theory has persisted in stellar spectroscopy, even if it requires some semi-empirical adjustment.  Over recent decades, the so-called Anstee-Barklem-O'Mara (ABO) theory has offerred a viable alternative.  This theory is a development of the theory by \cite{brueckner_collision_1971}, in which interaction potentials are calculated without resort to a multipole expansion, but via perturbation theory.  The Uns\"old approximation is applied in the case of neutral absorbers, and Coulomb wavefunctions employed, such that it retains the characteristic of the LFU theory of providing results that are independent of the element involved, and results can be tabulated with effective principal quantum number for different combinations of orbital angular momentum quantum numbers involved in the transition. \cite{brueckner_collision_1971} attempted to calculate the potentials numerically, which led to large numerical errors.  The ABO theory for neutrals has been developed in \cite{anstee_investigation_1991, Anstee1992, Anstee1995, Barklem1997a, Barklem1998b}, and a main advance was the advent of symbolic computing, thus allowing the majority of the required integrations in the potential calculations to be performed analytically.  The theory further makes improvements regarding the treatment of the collision dynamics, in particular accounting for rotation of the internuclear axis \citep{Roueff1974}, the neglect of which can lead to differences of order 30\%.  The ABO theory has been seen to be remarkably successful on the basis of comparison with solar spectra assuming abundances derived by other means \citep{Anstee1995, Barklem1997a, Barklem1998b} and with other more detailed calculations (see tables~\ref{tab:broad_na} and~\ref{tab:broad_ca}).   This agreement is somewhat surprising given the approximations of the ABO theory, such as the Uns\"old approximation and neglect of spin and ionic effects.  The ABO theory has also been successfully extended to absorbing ions, but in this case part of the Uns\"old approximation is not valid, and this leads to results that are dependent on the ion involved \citep{Barklem1998a,Barklem2000d}.  Data for many lines have been calculated explicitly \citep{Barklem2000c} and made available through the Vienna Atomic Line Database \citep[VALD; see][and references therein]{ryabchikova_major_2015}.  Code to interpolate in the tabulated data for neutrals has also been made available \citep{Barklem1998c,barklem_abocross_2015}. Large-scale calculations have been performed for \ion{Fe}{ii} \citep{Barklem2005c}, as well as for \ion{Cr}{ii} and \ion{Ti}{ii} (unpublished).  The results for \ion{Fe}{ii} and \ion{Cr}{ii} have been made available through VALD, and the results for \ion{Ti}{ii} should be made available in the near future.  An example of the improvement in using ABO theory compared to LFU theory for fitting the \ion{Ca}{ii} infrared triplet in the solar spectrum is shown in fig.~\ref{fig:abo}.

Line shifts are also reported in tables~\ref{tab:broad_na} and~\ref{tab:broad_ca}, where available.  These are unlikely to be of importance in any astrophysical context \citep[see][]{allendeprieto_limited_1997}.  However, it is interesting to note that the theoretical results show a large spread, in the case of Ca $4p$--$5s$ not even agreeing on the sign of the shift.  This is explained by the fact that the line shift is very sensitive to close collisions, which depend on the accuracy of the interaction potentials at very short internuclear distance.

Finally, a major limitation of the ABO calculations is that they do not cover transitions involving Rydberg states; these are often important in the study of lines in the infra-red.  This is primarily due to fact that inelastic processes, which are not in the ABO model (or the LFU model), become important and even dominant.   Such data may be calculated in the impulse approximation, see for example \cite{HoangBinh1995,osorio_mg_2015}.  It is unclear what the accuracy of such calculations is, but reasonable results are obtained in astrophysical modelling \citep{osorio_mg_2015}.

\subsection{Depolarisation of spectral lines}
\label{sec:depol}

Observations of linear polarization at the limb of the Sun, the so-called ``second solar spectrum'', provide an important window to solar physics, in particular the magnetic field of the quiet Sun \citep[e.g.][]{stenflo_second_1997}.  The polarization is reduced by magnetic fields, via the Hanle effect, as well as by collisions.  Thus, to interpret such spectra quantitatively requires non-LTE radiative transfer modelling including depolarisation of spectral lines by collisions.  

Though not directly relevant to abundance analysis, depolarisation of spectral lines by collisions has very similar physics to the broadening of spectral lines, and thus should be mentioned in this context.    Many of the approximations used are exactly the same. The rates for both processes depend on scattering matrix elements between the Zeeman sublevels of the relevant levels, and these matrix elements may be obtained in exactly the same manner in both cases, by solving the coupled equations for a given set of interaction potentials.  However, the equations for the depolarisation and broadening rates differ significantly, as they involve rather different linear combinations of the relevant matrix elements.  The similarities and differences between the theories for the two processes has been examined in detail recently by \cite{sahal-brechot_collisional_2014}.  A large amount of work has been done by 
\cite{Derouich2003b, Derouich2003a, Derouich2004, Derouich2004a, Derouich2005, Derouich2005a, Derouich2007, Sahal-Brechot2007} to adapt the ABO broadening theory to collisional depolarisation for both simple and complex atoms and ions. The similarities in the basic theory have led a number of studies to attempt to find simple analytical relationships between the depolarisation and broadening rates \citep{Faurobert-Scholl1995,Smitha2014a}, in order to be able to estimate depolarisation rates not otherwise available.  However, \citeauthor{sahal-brechot_collisional_2014} conclude that this is not possible, at least analytically, for the theories presently in use, due to the very different way matrix elements are combined.  Recently, \cite{derouich_unified_2015} have presented numerical relationships between the DSB depolarisation rates and the ABO broadening rates, which can be used to estimate depolarisation rates, given broadening data.

\subsection{Inelastic collisions}
\label{sec:inelastic}

Modelling stellar atmospheres in non-LTE requires information on all important inelastic processes occurring to the atom or ion of interest, and this presents a major stumbling block to such modelling.  First one must decide, among the numerous possibilities, which perturbing particles are important enough to need to be accounted for, and secondly calculate collision rates for all such processes.  In FGK stellar atmospheres, the usual candidates are electrons, since they are light and fast moving, and hydrogen atoms, since they are the most abundant particle.

\begin{figure}
   \centering
   \includegraphics[angle=0,width=1.0\textwidth]{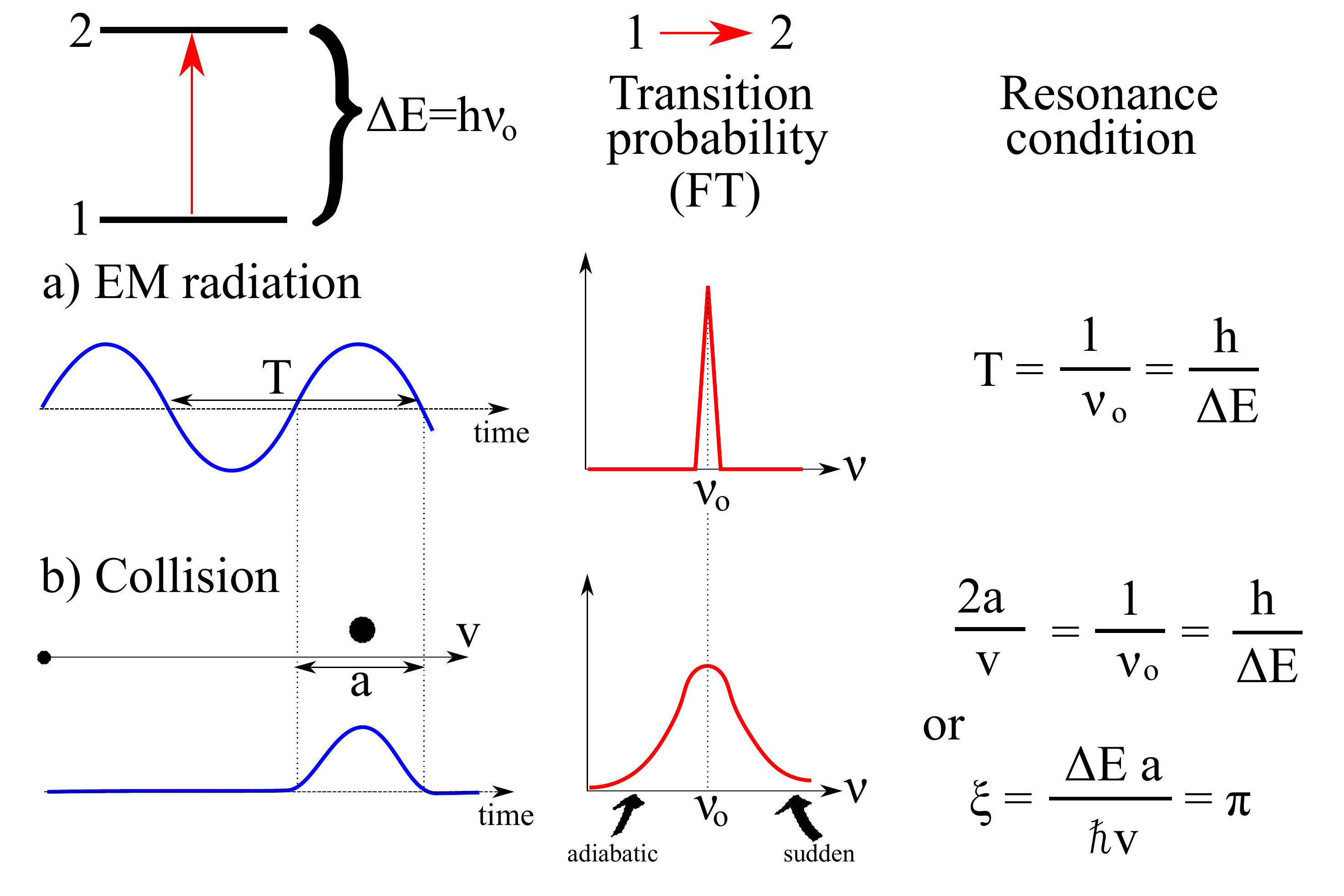} 
   \caption{Explanation of the Massey criterion for collisional excitation by analogy with excitation by single-frequency electromagnetic (EM) radiation, adapted from \cite{Andersen1981}.  In the case of single-frequency EM radiation (such as a laser), the transition resonance occurs when the frequency of the field $\nu$ matches the natural frequency of the transition $\nu_0$.  The transition probability as a function of $\nu$ is estimated by the Fourier transform (FT) of this field, and gives the well-known sharp peak.  In the case of a collision, by analogy, transition resonance occurs when the characteristic interaction length $a$ and the collision velocity $\mathrm{v}$ combine to produce a field with a characteristic timescale $T = 1/\nu = 2a/\mathrm{v}$ corresponding to the inverse of the natural frequency.  As the field has a broader frequency spectrum, the transition probability spectrum is also correspondingly broader as a function of the characteristic frequency of the field caused by the collision.  }
   \label{fig:massey}
\end{figure}

The Massey criterion, derived by \cite{Massey1949} from consideration of experimental and theoretical predictions, provides a rather simple way to estimate if a collision process might be efficient (i.e. near resonance).  The Massey parameter $\xi$ is the ratio of the timescale that characterises the rate of change of electronic wavefunction due to nuclear motion $\tau_\mathrm{nuc}$ and the relevant internal timescales of electronic motion associated with the transition of interest $\tau_\mathrm{el}$,
\begin{equation}
\xi = \frac{\tau_{nuc}}{\tau_{el}}. 
\end{equation}
The timescale characterising the electronic motion is given by the inverse transition frequency
\begin{equation}
\tau_{el} \sim \frac{1}{\Delta \omega} = \frac{\hbar}{\Delta E},
\end{equation}
where $\Delta \omega$ is the frequency of the transition of interest, and thus $\Delta E$ is the energy difference between states.  The timescale of nuclear motion can be estimated by
\begin{equation}
\tau_{nuc} \sim \frac{a}{\mathrm{v}},
\end{equation}
where $a$ represents the important length scales (i.e. it is an estimate of the important impact parameters), and $\mathrm{v}$ is the collision velocity.  Thus
\begin{equation}
\xi = \frac{\Delta E}{\hbar} \frac{a}{\mathrm{v}}. 
\end{equation}
For $\xi \gg 1$ we have adiabatic conditions, i.e. the timescales of nuclear motion are large compared to those of electronic motion, and the electron cloud has time to adjust to the slow nuclear motion.  In this regime transitions are very unlikely.  For $\xi \sim 1$ we have resonant conditions and transitions are  likely.  For $\xi \ll 1$ we have sudden conditions i.e. the timescales of nuclear motion are small compared to those of electronic motion, and the system has little time to react to the collision.  In this regime transitions are generally still probable, but not as likely as at resonance.  The physical explanation of these main characteristics can be understood by analogy to the radiative case and is demonstrated in fig.~\ref{fig:massey} following \cite{Andersen1981}.    

Such considerations were applied to discussion regarding the candidate perturbers for causing inelastic processes in stellar atmospheres, notably \cite{plaskett_interpretation_1955}, and lead to the conclusion that for typical optical transitions in the conditions of FGK stellar atmospheres electron collisions are likely to be in the resonant or sudden regime, while heavy particles collisions such as those with hydrogen are likely to be in the adiabatic regime  (see \cite{Lambert1993} for a full explanation of the argument).   This, factored together with the higher thermal velocity of electrons, has led to the general ``handwaving'' conclusion that electrons are likely to be the most important perturber as regards inelastic processes in FGK stellar atmospheres.  However, even if heavy particle collisions are expected to have small cross sections, and their lower relative speed to lower rate coefficients, this may be overcome by the far greater abundance of heavy particles, particularly hydrogen atoms.  In the line forming regions of FGK stellar atmospheres, hydrogen atoms typically outnumber electrons by four orders of magnitude in stars of solar metallicity.  In metal-poor stars the disparity is even greater.  Thus it is very important to know the efficiency of these collisional processes quantitatively in order to be able to accurately assess if they are important or not.   

Detailed calculations of inelastic collision processes are, unfortunately, always very difficult as they involve intricate atomic or molecular structure calculations to produce wavefunctions and energies required to perform quantum scattering calculations.  Scattering calculations for atoms always require approximations as exact solutions are not available.  Generally, the problem of electron collisions is far more tractable than that of heavy particle collisions, as the former relies on atomic structure calculations, while the latter requires molecular structure calculations (of the quasi-molecule formed during the collision) with more degrees of freedom, in particular internuclear distance.  Further, the nuclear motion and its relationship to the electronic motion adds an additional complication in solving the coupled channel equations.   Some insight into why the heavy particle collision problem is intrinsically more difficult is given by considering a more rigorous formulation of the Massey parameter in terms of the relevant molecular states during such a collision.  Such theoretical analysis indicates that the Massey parameter is more appropriately written as \citep[e.g.][]{nikitin_selected_1978,Nikitin1984,Nikitin2006}
\begin{equation}
\xi = \frac{\Delta U}{\hbar} \frac{a}{\mathrm{v}}, 
\label{eq:massey}
\end{equation}
where $\Delta U$ is now the splitting between adiabatic potential energy surfaces.   Thus, a more correct calculation even for the simple Massey parameter requires detailed knowledge of the potential energy surfaces of the system, which requires complex quantum chemistry calculations of the quasi-molecule formed during the collision for states corresponding asymptotically to both ground and excited states of the atoms of interest.  It has, however, been understood for some time \citep[e.g.][]{Bates1962}, that in collisions between atoms, adiabatic potentials approach each other, so-called ``pseudo-crossings'' occur, and thus $\Delta U \ll \Delta E$ leading to the possibility that transition probabilities may become large even for slow collisions.  The effect of the nuclear dynamics is that the electronic structure of the atom-atom system during the collision can differ significantly from the electronic structure of the separated atoms, leading to interactions that are very different in character to those found in the separated atoms. 

In the following section, the progress that has been made on understanding and calculating inelastic processes due to impacts by electrons and hydrogen atoms, and their relationship to modelling FGK stellar spectra, will be reviewed. 

\subsubsection{Electrons}
 
The $R$-matrix approach, discussed in \S~\ref{sec:photo-ion}, has again been of particular importance as a route to obtaining electron-atom (ion) excitation cross sections.  Again, for a review of the basic method and many of the early calculations, see \cite{Burke1993}.  An equally important aspect for astrophysical applications has been that the codes for calculation with the $R$-matrix method have been made freely available.  Various codes are available for the initial atomic structure calculations to calculate required orbitals, such as \texttt{CIV3} \citep{Hibbert1975}, \texttt{SUPERSTRUCTURE} \citep{eissner_techniques_1974}, and \texttt{AUTOSTRUCTURE} \citep{badnell_dielectric_1986}.  Code for the solving the internal region problem, \texttt{RMATRX I}, has been provided by \cite{Berrington1995}.  These codes are used for all electron-ion-photon interaction processes, with additional codes required depending on which process is to be calculated.  It is a particularly appealing feature of the $R$-matrix method, that the $R$-matrix is given at all energies by a relatively simple expression, so that once the internal region problem is calculated, the scattering quantities can be computed for a large range of processes and for a large number of energy values at quite low computational expense.  In the case of electron impact excitation, various codes are available for solving the external region problem and obtaining the required cross sections, such as \texttt{FARM} \citep{burke_farm_1995} and \texttt{STGF} \citep{berrington_atomic_1987}, which has been modified for treating collisions with neutral atoms by \cite{Badnell1999}.
 
The above standard $R$-matrix approach is based on atomic structure calculations using the configuration interaction method employing orthogonal Slater-type orbitals.  A single set of orbitals is optimised on the atomic energy levels by means of the variational principle.  This approach has two main problems, namely that it cannot account for a) the term dependence of the physical orbitals, or b) the coupling to the ionisation continuum.   In the standard approach, this problem is often addressed by employing ``pseudo-orbitals'' and ``pseudo-states'' (states constructed with at least one such orbital in a dominant configuration).  The problem of inclusion of the coupling to the ionisation continuum is particularly important in neutral atoms, as the continuum contributes significantly to the atomic polarisability and thus the long-range interaction potential, as well as providing a channel for ionization processes themselves.  For example, the inclusion of pseudo-states has been shown by \cite{Griffin2001} \citep[see also][]{Osorio2011} to be important at intermediate energies in the case of electron collisions with Li.  This method, the so-called $R$-matrix with pseudo-states (RMPS) approach, has been exploited recently by \cite{Osorio2011, osorio_mg_2015} to calculate cases of particular interest for FGK stellar spectra, namely electron collisions on Li and Mg, while calculations have been performed by \cite{Barklem2007a} without pseudo-states.  In the cases of Li and Mg pseudo-orbitals have been optimised on the polarisability \citep{ledourneuf1976, hibbert_atomic_1977,plummer_lowenergy_2004}.  Calculations have also been performed by \cite{Barklem2007a} for O, but without pseudo-states; the influence of the continuum is expected to be smaller due to the large ionisation energy of \ion{O}{i}, particularly for low-energy collisions of most interest in FGK stellar atmospheres.  Note that the coupling to the ionisation continuum plays its most important role at intermediate energies, above the ionisation threshold.  Having said that, better calculations for this important case would be worthwhile.
 
An important advance during the last 15 years or so has been the $B$-spline $R$-matrix (BSR) method.  This method employs non-orthogonal $B$-spline orbitals, rather than the more usual case of orthogonal Slater-type orbitals, optimized on each individual term, instead of simultaneously optimised on all states of the atom.  This permits a compact configuration interaction expansion that is more flexible and thus more accurate.  The main advantage of this approach is that it removes the need for pseudo-orbitals to correct for problems resulting from the limited single set of orbitals used in the standard approach.  This drastically reduces the problem of pseudo-resonances occurring due to unphysical thresholds resulting from the appearance of pseudo-states (states with large contributions from pseudo-orbitals).  The method has the downside that it is more complicated and computationally expensive than the standard approach; however, codes to do such calculations are freely available \citep{zatsarinny_bsr_2006}.  For a recent review of the method and the calculations performed with it, see \cite{zatsarinny_b_2013}.  In particular, we note that calculations have been performed for over 20 neutral atoms, many of which are of interest in FGK stellar atmospheres.  Here, often there is a problem as regards the publication of the data required: rate coefficients for transitions between as many states as possible, and including excited states, are needed, preferably available in electronic form.  We note, however, that this problem is being addressed.  For example calculations of electron collisions with Mg were performed by \cite{zatsarinny_cross_2009}, and recently tabulated data suitable for non-LTE applications in stellar atmospheres has appeared \citep{merle_effective_2015}.

Finally, we mention another advanced close-coupling method, the convergent close-coupling method \citep[CCC][]{Bray1992,Bray1995}.  This method employs orthogonal Laguerre basis functions.  The main advantage of the method is that it can be demonstrated, at least for one-electron excitations, that scattering amplitudes converge with increasing basis size (with the effects of pseudo-resonances diminishing).  However, the method has the disadvantage of requiring considerable computational effort.  It has been used for hydrogen-like atoms \citep{bray_convergent_1994-1, bray_convergent_1994}, quasi two-electron targets (e.g. alkaline-earth metals) as well as noble gases \citep[e.g.][]{fursa_convergent_1997,bray_electrons_2002,bray_benchmark_2011,bray_electron-_2012}.  At present the only case of interest for FGK stellar atmospheres where an extensive set of rate coefficient data for excitation is available from the CCC method is that for Li \citep{Schweinzer1999}.  More such data sets for other targets would be of significant importance.

The results of the above mentioned modern close-coupling methods, RMPS, BSR, and CCC, all generally agree well with experiment, with agreement being best with simple atoms and worst with complex atoms, as would reasonably be expected.  \cite{bartschat_closecoupling_2015} have, based on the agreement with experiments, estimated the relevant uncertainties in collision cross sections for the processes that have a significant effect on plasma modelling.  For the cases relevant to FGK atmospheres and their spectra, they estimate 3\% uncertainty for hydrogen, 10\% for the light alkalis and quasi-two electron systems (Li, Be, Na, Mg, K, Ca), 20\% for the heavy alkalis (Rb, Cs) and the light open-shell atoms (C, N, O, Al, Si, etc.) and 30\% for the heavy quasi-two electron systems (Sr, Ba).  For heavy open-shell systems, such as the iron-group elements, there are very few calculations for both neutral and singly ionised targets, examples being \ion{Fe}{ii} \citep{nussbaumer_atomic_1980, ramsbottom_electron_2002, zatsarinny_benchmark_2005, ramsbottom_electronimpact_2007}, \ion{Cu}{i} \citep{zatsarinny_electronimpact_2010,zatsarinny_electron_2010} and \ion{Zn}{i} \citep{zatsarinny_benchmark_2005-1}.  These calculations only cover a limited number of low-lying states, and do not generally have the completeness necessary for non-LTE modelling in stellar atmospheres.  Calculations of electron-impact excitation of iron-group elements in neutral and singly ionised states, with relatively complete coverage are an important goal for the study of stellar atmospheres.  The impressive progress in this field over the last 25 years should be a basis for optimism on this front.

In the absence of modern close-coupling calculations, models are forced to resort to approximate methods, the most usual of which are the formulae of \cite{vanRegemorter1962} and \cite{Seaton1962b}.  Both these formulae are based on the Bethe approximation, in which the cross section for allowed transitions is written in terms of the optical oscillator strength.  \cite{Seaton1962} has shown that this may be interpreted as the colliding electron making a free-free transition in the field of the atom, which is subsequently absorbed by the atom to produce the excitation.  The \cite{vanRegemorter1962} formula is a semi-empirical one based on the Born approximation, which is well known to overestimate cross sections at low energy. The \cite{Seaton1962b} formulae employ the semi-classical impact parameter method for neutrals, and have been extended to the case of ions by \cite{Burgess1976}.  The impact parameter generally gives much better results than the van Regemorter formula, as would be expected since the former does not make the Born approximation; see discussion in \cite{Bely1970}.  The superiority of the impact parameter method also seems to be borne out in astrophysical modelling, with \cite{osorio_mg_2015} recently showing that it provides much better modelling of the solar \ion{Mg}{i} infrared emission lines, which probe Rydberg levels not easily covered by close-coupling calculations, than if the van Regemorter formula is used.  

Thus, it is clear that where possible modern close-coupling calculations should be done and used in non-LTE modelling.  Where these are unavailable, for example transitions involving Rydberg states, then the impact parameter method should be employed for optically allowed transitions.  For forbidden transitions, there is no clear path to estimate collision rates.  A common approach is to estimate the collision strength $\Omega_{ij}$ based on close-coupling calculations for other atoms or for forbidden transitions in the same atom.  Recent work by \cite{osorio_mg_2015} suggests an approach where the $\Upsilon_{ij}/g_i$ is estimated from other transitions, where $\Upsilon_{ij}$ is the thermally-averaged collision strength and $g_i$ is the statistical weight of the initial state.  It is also suggested that exchange and non-exchange transitions, transitions where the spin is changed or unchanged, respectively, should be treated separately, and it was found that the behaviour of these two groups was indeed significantly different in the case of \ion{Mg}{i}.  It would of course be more satisfactory to have a physically-based approximate formula for estimating collision rates in the case of forbidden transitions; however, this seems difficult as even the simplest estimates for forbidden transitions require knowledge of wavefunctions (for example in the Born or Ockhur approximations), which is avoided in the allowed transition case as this information essentially enters via the oscillator strength.  Thus it is unclear that any approximate approach would have significant advantages over doing coupled-channel calculations.

\subsubsection{Hydrogen} 
\label{sect:hinelast}

In the case of slow heavy-particle collisions, where nuclear motion is much slower than electronic motion, expansion of the electronic wavefunction for the colliding system in terms of wavefunctions of the isolated atoms, as is done for electron collisions and high-energy heavy-particle collisions, is unsuitable.  The expansion should be in terms of the wavefunctions of the quasi-molecule formed by the target and the perturber \citep[e.g.][]{mott_theory_1931,mott_theory_1949,bates_inelastic_1953,Bates1962}; see also \citet[][Sect.~2.2]{Barklem2011} for a comparison of quantum collision theory for various cases.  As discussed above, even a simple estimate of the Massey parameter for hydrogen atoms colliding with another atom, requires detailed knowledge of the (adiabatic) potential energy surfaces for the quasi-molecule formed during the collision.    

A complete treatment requires the solution of the coupled-channel Schr\"odinger equations, for which various couplings between nuclear and electronic motion need to be known in addition to potential energies.  In the standard adiabatic Born-Oppenheimer approach these are the radial and rotational (or Coriolis) couplings, while in the diabatic approach these take the form of the off-diagonal matrix elements of the Hamiltonian operator (the diagonal elements being the potential energies in this representation).   Even for simple cases such as alkali-hydrides, potential energy surfaces for (the very lowest) excited states were not available until the 1970's \citep[e.g.][]{docken_lih_1972, sachs_mcscf_1975, meyer_pnoci_1975, olson_interaction_1980, partridge_theoretical_1981, mo_calculation_1985}.   Couplings have very rarely been calculated \citep[e.g.][]{mo_calculation_1985, Gadea1993,Guitou2010}.  Further, and perhaps even more importantly, the so-called ``electron translation problem'' \citep{bates_electron_1958} in the standard adiabatic Born-Oppenheimer approach, whereby the choice of origin of electronic coordinates leads to non-zero asymptotic couplings and seemingly to ambiguity in the solution of the coupled-channel equations, has only recently found a satisfactory solution \citep{belyaev_dependence_2002}.  This work puts the standard Born-Oppenheimer approach on a solid foundation, proving that the solution is independent of the choice of electronic coordinates, and, if treated correctly provides the correct solution to the coupled-channel Schr\"odinger equations.   A major advance has been the development of the reprojection method \citep{belyaev_revised_2010, grosser_approach_1999, belyaev_electron_2001,Belyaev2009}, which allows solution without resort to ``electron translation factors'' \citep[e.g.][]{errea_nonadiabatic_1986}.  

\begin{figure}
\center
\includegraphics[width=1.0\textwidth]{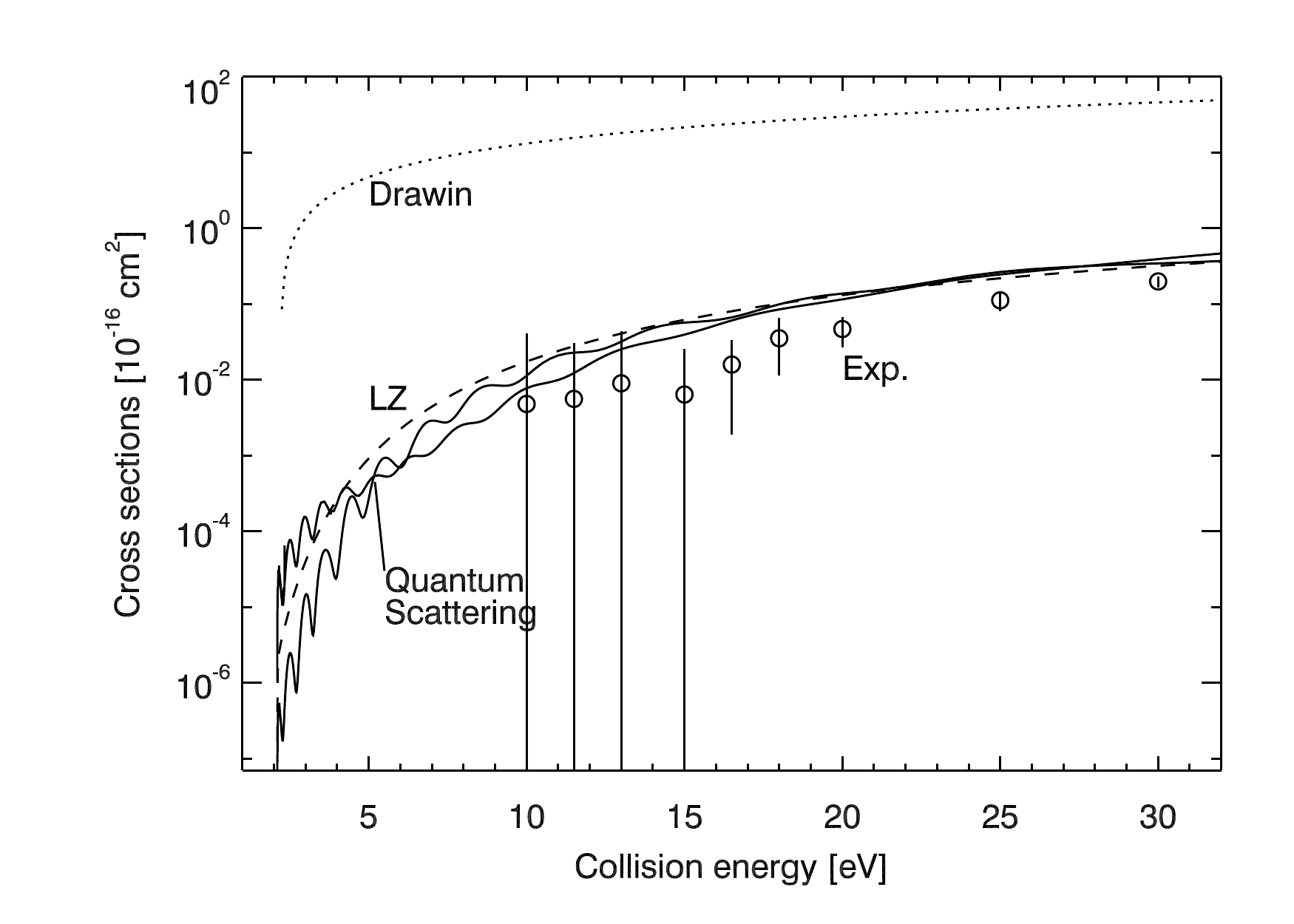}
\caption{Comparison of cross sections for the process Na($3s$) + H $\rightarrow$ Na($3p$) + H. The dotted line shows the Drawin cross section and the dashed line the Landau-Zener cross section. The full lines are two quantum scattering calculations using different input quantum-chemical data (MRDCI, which is largest near the threshold, and pseudo-potential; see \cite{Belyaev2010}, for details). The circles show the experimental data of \cite{Fleck1991} \citep[see also][]{Belyaev1999} with 1$\sigma$ error bars.  Credit: \citeauthor{Barklem2011}, A\&A, 530, A94, 2011, reproduced with permission \textcopyright\ ESO.}
\label{fig:na3s_to_3p_cross}     
\end{figure}

This approach has been applied to various cases of low-energy collisions of hydrogen with targets of interest to stellar astrophysics, namely Li \citep{Belyaev2003}, Na \citep{Belyaev1999,Belyaev2010} and Mg \citep{Guitou2011,Belyaev2012}, based on quantum chemistry calculations of potentials and couplings predominantly from \cite{Croft1999a}, \cite{dickinson_initio_1999} and \cite{Guitou2010}, respectively.  The calculations agree well with the extremely limited experimental data available.  For excitation due to hydrogen atom impact there is one experiment at low to intermediate energy, 15--1500 eV, for Na($3s$) + H $\rightarrow$ Na($3p$) + H \citep{Fleck1991}.  Comparison of theoretical results mentioned above with the experimental results is shown in fig.~\ref{fig:na3s_to_3p_cross}.  Note, the experimental results do not go down to threshold, where at present we only have theoretical predictions.  The near-threshold cross sections, which determine the collision rate at the temperatures of interest for FGK stellar atmospheres since $kT\sim 0.2$--0.6~eV and thus the threshold is on the tail of the Maxwellian velocity distribution, are seen to be sensitive to the input quantum chemistry, giving cross sections differing by two orders of magnitude at threshold.  This turns out to be of no astrophysical consequence as will be discussed further below.  Rate coefficients suitable for astrophysical applications based on these calculations have been presented in \cite{Barklem2003b,Barklem2010,Barklem2012} and applied to non-LTE modelling of FGK stellar spectra in \cite{Barklem2003b}, \cite{Lind2009b}, \cite{Lind2011}, and \cite{osorio_mg_2015}.   

\begin{figure}
\center
\includegraphics[width=1.0\textwidth]{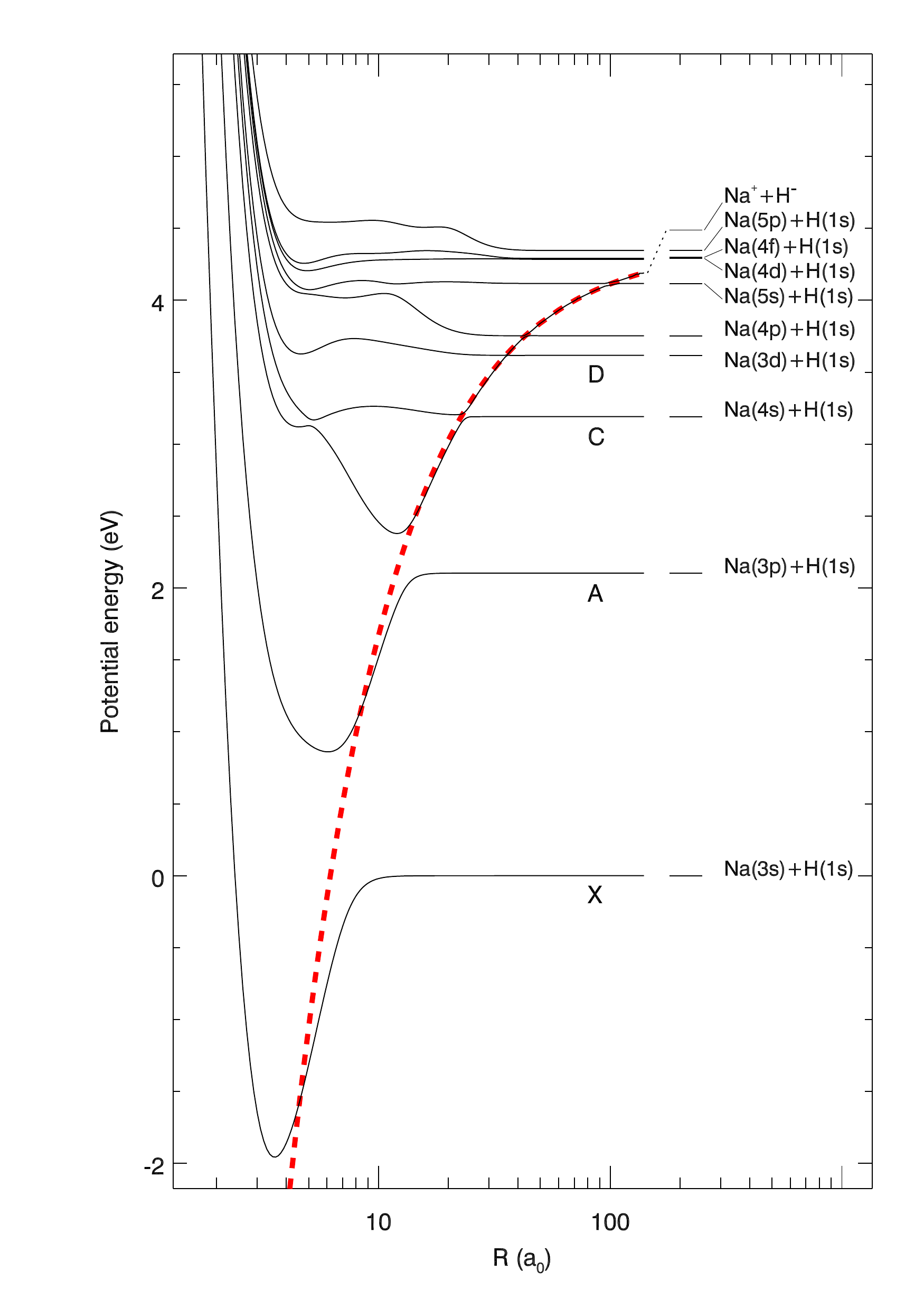}
\caption{Adiabatic potential energies as a function of internuclear distance $R$ for the lowest ten $^1\Sigma^+$ states of the NaH quasi-molecule from pseudo-potential calculations by  \cite{dickinson_initio_1999}. The atomic states and energies at dissociation are shown at the right-hand side of the figure. The thick, dashed (red) line shows the $1/R$ interaction corresponding to the Coulomb interaction for the pure ionic configuration Na$^+$+H$^-$ at long range. Credit: \citeauthor{Barklem2011}, A\&A, 530, A94, 2011, reproduced with permission \textcopyright\ ESO. }
\label{fig:nah_crossings}     
\end{figure}

\begin{figure}
\center
\includegraphics[width=1.0\textwidth]{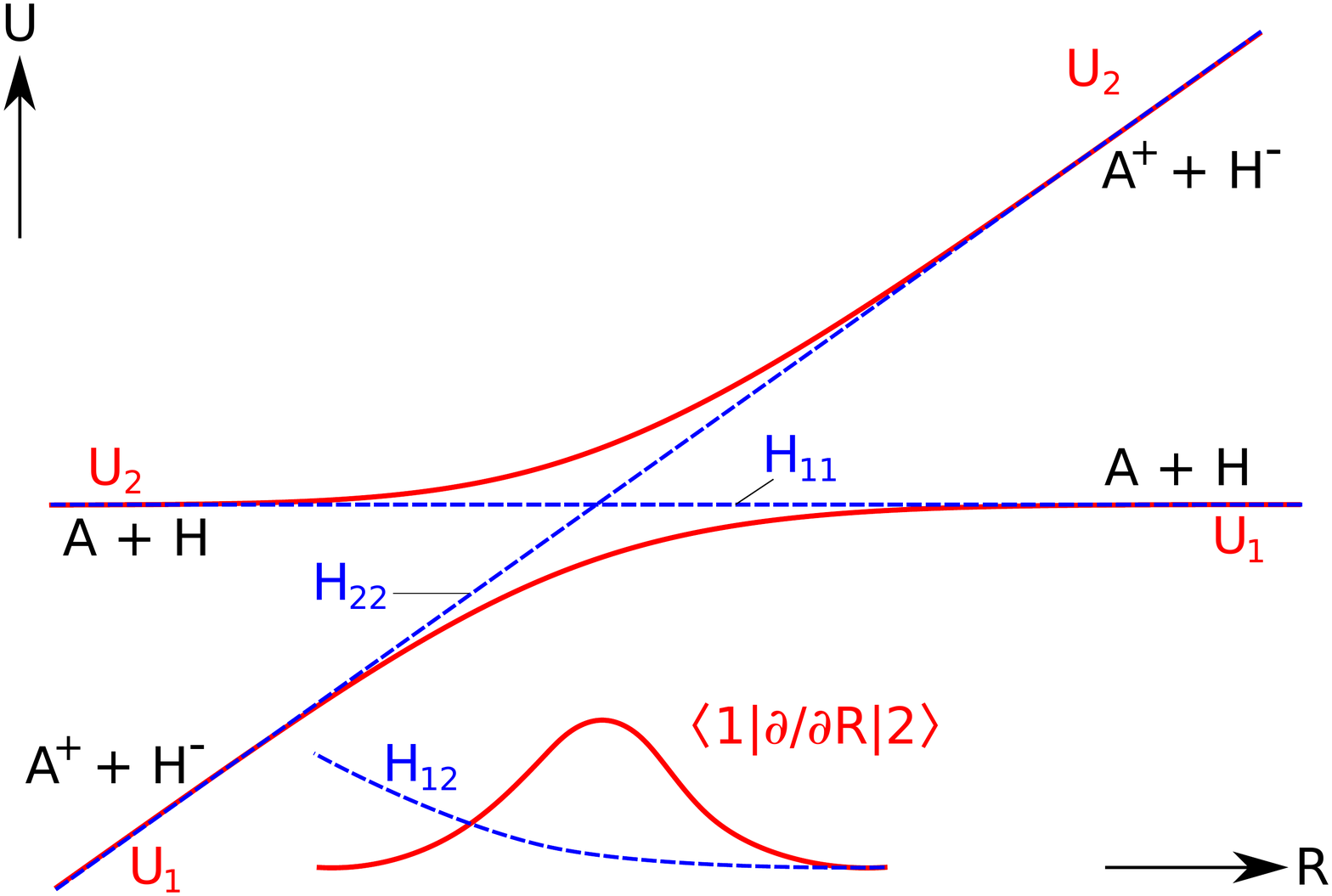}
\caption{Illustration of behaviour of potentials curves and couplings at an avoided ionic crossing, as a function of internuclear distance $R$.   The (red) full curves show the potential curves in the adiabatic representation, $U_1$ and $U_2$, as well as the non-adiabatic radial coupling $\langle 1 | \partial / \partial R | 2 \rangle$.  At the crossing, the adiabatic wavefunctions change character from covalent to ionic configurations, and vice versa.  The (blue) dashed curves show the energies in the diabatic representation, $H_{11}$ and $H_{22}$, and $H_{12}$ is the coupling matrix element. Adapted from fig.~23 of \cite{Bates1962} and figs. 5.5 and 5.7 of \cite{nikitin_selected_1978}.  See  Sect.~5.3 of \cite{nikitin_selected_1978} for a more complete description.}
\label{fig:crossing}     
\end{figure}

These calculations have highlighted the importance of pseudo-crossings associated with avoided ionic crossings, a mechanism which has been understood for some time \citep[e.g.][]{Bates1962}, but due to the requirement of detailed quantum chemistry calculations, accurate calculations have been difficult.  That this mechanism was at work in inelastic collisions involving hydrogen atoms was flagged by \cite{Fleck1991}.  Comparisons with Landau-Zener model calculations in that work, as well as later comparisons of revised experimental data with detailed quantum scattering calculations by \cite{Belyaev1999} demonstrated that the experimental data could be reasonably explained by non-adiabatic transitions associated with pseudo-crossings in the NaH molecular potentials, particularly those associated with interaction of the covalent Na+H and ionic Na$^+$+H$^-$ configurations, which are thus called avoided ionic crossings.   The adiabatic potentials for the Na+H quasi-molecule with these avoided ionic crossings are shown in fig.~\ref{fig:nah_crossings}, where the influence of the ionic configuration in perturbing the potentials from their behaviour at large internuclear distance is clearly seen.  Similar effects are seen in other quasi-molecules, for example LiH \citep{Croft1999a} and MgH \citep{Guitou2010}.  The physical meaning of this behaviour is that, in a quasi-molecule A+H, starting at long range from two atoms in a covalent configuration where the electrons are associated to their respective nuclei, as the atoms are brought closer together they come to an internuclear distance where this covalent configuration has the same energy as the ionic configuration A$^+$+H$^-$.  In the adiabatic representation, if the corresponding states have the same symmetry, the potential curves avoid each other according to the von Neumann-Wigner non-crossing rule, and the molecular state corresponding to the covalent configuration at long range, changes to have a predominantly ionic configuration at shorter internuclear distances.   Thus, the adiabatic wavefunction of the system changes dramatically in the crossing region leading to large non-adiabatic couplings, particularly the radial coupling matrix element $\langle 1 | \partial / \partial R | 2 \rangle$, where $|1\rangle$ and $|2\rangle$ are the adiabatic wavefunctions, which describes the coupling between nuclear motion in the direction of the internuclear axis and changes in the electronic wavefunctions.   The behaviour of the potentials and couplings in the adiabatic and diabatic representations is illustrated in fig.~\ref{fig:crossing}.  For a more thorough discussion of the theory of slow atomic collisions and precise definitions of terms, see for example \cite{nikitin_selected_1978}.

Thus the physical interpretation of this mechanism for non-adiabatic transitions is that during a collision, at avoided ionic crossings, the valence electron associated with the atom A has a probability to tunnel to the H atom resulting in a predominantly ionic charge distribution. Later during the collision at a different avoided crossing, there is a probability that the electron may tunnel back to a different covalent molecular state leading finally to a different final state of the atom A, and excitation or deexcitation A($nl$) + H $\rightarrow$ A($n'l'$) + H. The electron may also stay with the hydrogen atom leading to ion-pair production A + H $\rightarrow$ A$^+$ + H$^-$. Calculations for Li+H and Na+H mentioned above have generally shown that even the largest cross sections for excitation are typically small compared with those for ion-pair production from certain states. This can be understood in terms of that a large cross section for an excitation process results from passing two avoided ionic crossings, one of them higher-lying and passed with a small probability of leading to a covalent state due to small adiabatic energy splittings at highly excited crossings.  However, for an ion-pair production process the system passes this same crossing with a large probability to remain in the ionic state.  Calculations have shown that cross sections for charge transfer processes (i.e. ion-pair production A + H $\rightarrow$ A$^+$ + H$^-$, and its inverse process mutual neutralisation) involving the first excited $S$-state are particularly large.  The reason for the large cross sections from these particular states can be understood by considering that owing to the superposition of two factors: first, that with increasing internuclear distance of the crossing region and thus the maximum possible cross section increases, while secondly the coupling between adiabatic states decreases with increasing internuclear distance.  This second factor may be understood simply in terms of that the electron has a wider barrier to tunnel through.  These competing factors lead to a maximum in ion-pair production cross sections for mildly excited states which have crossings with the ionic state at intermediate internuclear distances.  Among these states, $S$-states then generally show the largest cross sections due to the large statistical weight of the initial channel.

The astrophysical applications mentioned above have shown the importance of the charge transfer processes with large cross sections in modelling FGK star spectra.\footnote{On a historical note, it is worth to comment that charge transfer was first considered in astrophysics by \cite{Chamberlain1956} after \cite{Bates1954} noted the process $\mathrm{H} + \mathrm{O}^+ \rightleftharpoons \mathrm{H}^+ + \mathrm{O}$,
should, due to the similarity of the ionisation potentials of H and O, have large cross sections at low energies.  \citeauthor{Chamberlain1956}, and later \cite{Field1971}, considered this process in the ISM.  Later \cite{judge_constraints_1986} also considered this process in stellar atmospheres. } In the cases of Li and Na, charge transfer processes involving the first excited $S$-state were found to be the only important processes.  Note, that the two-order-of-magnitude uncertainty in the Na($3s$) + H $\rightarrow$ Na($3p$) + H near-threshold cross sections, and thus in the rates, has no astrophysical importance.  In either case the rates are far too low to have significant effects in FGK stellar atmospheres.  In metal-poor stars, effects of the charge transfer collisions on derived abundances may be as large as 0.2 dex (60\%).  Comparison of total mutual neutralisation cross sections with experiments by \cite{peart_merged_1994} for Li$^+$+D$^-$ for collision energies 0.6--300~eV finds the theoretical result about 20\% greater at the lowest measured energy.  There is a general trend of the disagreement becoming larger at lower energies, and we recall that the collision energies of importance here are 0.2--0.6~eV \citep[see][]{Croft1999a,Barklem2003b}.  Although the charge transfer rates are important, the astrophysical modelling does not seem very sensitive to errors in the rates.  In the cases of Li and Na, increasing or decreasing the charge transfer rates by a factor of two, has effects on derived abundances of less than 0.01~dex.  In the case of Mg, which has the added complication of multiple spin systems, the situation is more complicated with both charge transfer processes and excitation processes with large cross sections seen to be important in the non-LTE modelling.  However, the general effects of hydrogen collisions were of the same magnitude, as large as 0.2 dex.  A more detailed study of the sensitivity of modelling of Mg spectral lines to the various collision rates is currently in progress \citep{osorio_mg_2016}.  We find that uncertainties related to the atomic collision data are typically of order 0.01 dex or less, although for few stellar models in specific lines uncertainties can be as large as 0.03 dex.  As these errors of order 0.01~dex are less than or of the same order as typical non-LTE abundance corrections, we conclude that the present collisional data are of sufficient accuracy to extract Li, Na, Mg abundances from high quality spectra more reliably than from classical LTE analysis.

Due to the lack of calculations for inelastic hydrogen collision processes, plasma modellers and astrophysicists have often resorted to simple estimates such as the ``Drawin formula'', a modified classical Thomson model approach \citep[see][]{Drawin1968,Drawin1969,Drawin1973,steenbock_statistical_1984,Lambert1993}.  The quantitative failing of this formula in the case of the excitation of Na is clearly seen in fig.~\ref{fig:na3s_to_3p_cross} \citep[see also][]{Lambert1993}, as well as in comparisons with the calculations discussed above \citep[see][]{Barklem2003b,Barklem2010,Barklem2012}, where discrepancies for optically allowed transitions range up to 6 orders of magnitude.  The Drawin formula cannot be used for forbidden transitions, which are often the majority of transitions in an atom, without some arbitrary assumption.  A major additional failing is that it cannot treat charge transfer processes, which as discussed have the largest cross sections and have been found to be the most important in astrophysical applications. The physical reasons for this failing have been discussed in detail by \cite{Barklem2011}.  

However, the fact that such an approach is still rather widely used, often with an astrophysical fudge factor, despite its clear failings \citep[e.g.][]{steffen_photospheric_2015} emphasises the need to solve this issue and obtain better data for many atoms of astrophysical interest.  The approach used for the calculations above, with detailed quantum chemistry calculations and full quantum scattering calculations is extremely labour intensive, as evidenced by the fact that calculations for three relatively simple systems have been done over the course of more than a decade, and required 10 man-years of work.   These methods are not currently tractable on reasonable timescales for quasi-molecules involving more complex atoms, especially those with open-shells such as O and Fe, which are of great astrophysical importance.  Thus, simpler methods to obtain estimates are clearly needed.   The detailed calculations and their astrophysical applications now provide a foundation for such work, as they provide an understanding of the physical mechanisms involved.   They are benchmark calculations against which to test approximations, and provide a guide to what accuracies are required for astrophysical modelling.  This work has now begun in earnest with some results already appearing for Al \citep{Belyaev2013a,Belyaev2013} and Si \citep{Belyaev2014}.  These particular calculations employ a model approach based on semi-empirical estimation of couplings at avoided ionic crossings \citep{Olson1971} and simplified description of the collision dynamics via the quantum branching-probability-current method \citep{belyaev_nonadiabatic_2011} and/or the multichannel model \citep{belyaev_1987,Belyaev1993,Belyaev2003}.  Both the branching-probability-current and multichannel methods are based on the Landau-Zener model of pseudo-crossings \citep{landau_1932,landau_1932-1,zener_nonadiabatic_1932}.  An alternate method using theoretical estimates of the potentials and couplings at avoided ionic crossings based on an improved version of an asymptotic two-electron model by \cite{Grice1974} and \cite{Adelman1977}, has also been developed \citep{barklem_excitation_2016}.  Comparisons with detailed calculations for Li, Na and Mg, indicate that at least for these simpler systems, such methods can provide rates accurate to within an order of magnitude for the partial processes\footnote{\emph{Partial} processes are those from one channel to a particular final channel, as distinct from \emph{total} processes from one channel to all possible final channels.} with the largest cross sections.  For processes with much smaller cross sections, the performance is considerably worse; however, these processes are almost certainly not astrophysically important.  Thus, these methods seem a productive route to obtaining reasonable estimates of charge transfer and excitation processes due to hydrogen collisions for the processes with the largest cross sections, which are likely to be the most important in modelling.  It will be important to test the sensitivity of the astrophysical modelling to uncertainties in the data, and then use this information to pinpoint where detailed calculations may be needed.  Future experiments on mutual neutralisation with the DESIREE facility \citep{thomas_desiree_2011} have the possibility to provide important checks on the charge transfer cross sections.  Astrophysical tests, especially centre-to-limb variation of spectral lines on the Sun also will provide important information about the general success or lack there-of of the non-LTE modelling, including the atomic collision data \citep[e.g.][]{AllendePrieto2004b,Pereira2009a}.
  
Finally, we note that the above addresses only low-lying states up to the A$^+$ + H$^-$ ionic limit 0.75~eV below the ionisation energy, but that for completeness modelling requires data also for Rydberg states.  This can be very important, as collisions among Rydberg states can provide a channel for populations to flow from ionised states to low-lying states of atoms, such as was found by \cite{Carlsson1992} in the modelling of \ion{Mg}{i}. \cite{hickman_theoretical_1983} has reviewed various theoretical approaches to low-energy collisions of Rydberg atoms with atoms and ions.  Recent work by \cite{osorio_mg_2015} has shown that the free-electron model within the impulse approximation, where the collision is considered as a binary interaction between the Rydberg electron and the perturbing atom, by \cite{Kaulakys1991} when used for \ion{Mg}{i} works reasonably well in reproducing the emission features.   This model is certainly to be preferred over the Drawin formula on physical grounds.   The expressions of \cite{Kaulakys1991} give the cross section for the process $nl\rightarrow n'l'$, but require the momentum-space wavefunctions of the initial state to be calculated.  Analytic expressions for the relevant integrals are given for $l=0$ in \cite{Kaulakys1985,Kaulakys1986}.  To calculate for other angular momentum states, the method of \cite{HoangBinh1997} may be used, which requires the calculation of Hankel transforms.  \citeauthor{HoangBinh1997} provide the relevant transforms only for the $l=0$ and 1 orbital angular momentum quantum numbers\footnote{Note there is also a misprint, see \cite{osorio_mg_2015}}.  I have derived the required Hankel transforms analytically using Mathematica for $l=0$--13, and provided IDL code  \texttt{MSWAVEF} for calculation of the wavefunctions by \cite{barklem_mswavef_2015}.  Code \texttt{KAULAKYS} to calculate cross sections and rate coefficients according to the free-electron model, based on these momentum-space wavefunctions is made available by \cite{barklem_kaulakys_2015}.

Often in non-LTE modelling, one wishes to resolve terms of different spin, yet models such as that of Kaulakys do not account for spin and give cross sections for $nl\rightarrow n'l'$.  Approximate formulae for the distribution of the $nl\rightarrow n'l'$ cross sections between different spin states can be derived in the scattering length (low-energy) approximation.  Consider a Rydberg atom with total electronic spin S, where the core electrons (i.e. all except the Rydberg electron) are assumed to have a total electronic spin $S_c$, which is assumed to be a good quantum number. The value of $S$ must be either $S_c-1/2$ or $S_c+1/2$, and assuming the core is not affected by the collision, as in the impulse approximation, the final total electronic spin $S'$ must also be either $S_c-1/2$ or $S_c+1/2$.  Generalising the calculation done in Sect 3.3.2 of \cite{osorio_mg_2015}, in the scattering length approximation, an estimate for the distribution of the cross section among spin states can be found:
\begin{eqnarray}
 \sigma (S \rightarrow S' ) & = & \frac{2S' +1}{2(2 S_c +1)} 0.392 \sigma_{nl\rightarrow n'l'}, \\
 \sigma (S \rightarrow S  ) & = & \left[ 1 - \frac{2S' +1}{2(2 S_c +1)} 0.392 \right] \sigma_{nl\rightarrow n'l'},
\end{eqnarray} 
where the scattering lengths for $e+$H scattering from \cite{Schwartz1961} have been used.

\section{Comments on molecules}
\label{sec:molecules}

The focus of this review is on FGK stars, and in particular on atomic data and processes relevant for the atmospheres of these stars.  The subject of cooler stars (M-type and later), where the visual spectrum is largely dominated by molecules, has been avoided.  A review of the role of molecules in the spectra of stars of K-type and cooler is given by \cite{bernath_molecular_2009}.  Nevertheless, a small amount of diatomic molecules is found in the atmospheres of FGK stars and their spectra, and can be unique and important probes of chemical composition.  Some elements, particularly the astrophysically very important cases of carbon, oxygen and nitrogen, often have few or no detectable atomic lines in the visible spectrum, yet produce molecular features that therefore become important diagnostics of the abundances of these elements.  Thus, it is worth commenting briefly on a few aspects here.  

First, it should be noted that there are a number of important cases in stellar atmospheres where adequate fundamental data are still unavailable.  A foremost example is NH, which is used to derive nitrogen abundances in metal-poor stars.  \cite{spite_first_2005} analysed the NH $A^3\Pi$--$X^3\Sigma$ band at 336 nm in a sample of metal-poor stars, and for a subsample also analysed the CN band at 389 nm.  They found a large systematic difference of 0.4~dex, which they attribute to uncertainties in the NH band data, namely the line positions, $gf$-values and dissociation energy.     

Second, non-LTE studies of molecular line formation are relatively rare.  Early studies such as those by \cite{thompson_conditions_1973} and \cite{hinkle_formation_1975} demonstrated that for cool stars the equilibrium for levels within the ground state of diatomic molecules, the rotational and vibrational levels, are collisionally dominated (noting that such levels are closely spaced, and thus the Massey criterion predicts collisions will be efficient).   Thus, LTE may be expected to be a good approximation for vibrational-rotational spectra of diatomic molecules.  The situation is far less clear for excited states, and non-LTE may need to be accounted for.  Work on non-LTE molecular line formation has to date predominantly focussed on CO \citep{thompson_conditions_1973, carbon_departures_1976, ayres_nonlte_1989, wiedemann_carbon_1994, berkner_3d_2015} and H$_2$O \citep{lambert_nlte_2013, ryde_systematic_2015}, due to their importance in cool stellar atmospheres.  Detailed calculations  require novel mathematical and computational methods for solving the radiative transfer problem \citep{schweitzer_nonlte_2000,lambert_new_2015}.  For atoms, the collisional data introduce a significant uncertainty in such modelling as well.

Abundance analyses using molecular features are exclusively carried out in LTE at present, and thus rely on equilibrium constants to estimate the relative abundances of molecules and their constituent atoms.  The compilations of \cite{Sauval1984} and \cite{Irwin1981} are still in wide use in the stellar atmosphere community.   The \citeauthor{Sauval1984} compilation is built on the data of \cite{Huber1979}.  Recently, \cite{barklem_partition_2016} have performed updated calculations of partition functions and equilibrium constants for around 300 diatomic molecules, using the data of \citeauthor{Huber1979} as a starting point, but updating data for around 90 molecules, in particular using data for ground states from \cite{Irikura2007, Irikura2009}.  These calculations also extended the temperature range of the calculations down to near absolute zero, and critically compiled dissociation energy data from the comprehensive compilation of experimental data from \cite{Yu-Ran2007}, and theoretical data from \cite{Curtiss1991}.  

Finally, we note that, as for atoms, the combination of laboratory measurements and astrophysical observations can also be used to enhance our knowledge of molecules.  For example, \cite{bernath_revised_2009} have revised molecular constants for OH using laboratory and solar spectra.

\section{Concluding remarks and outlook}
\label{sec:conclusions}

A recurring theme throughout the review, and going back to the development of Bohr's model of the hydrogen atom, is the interplay between theoretical atomic physics, laboratory measurements, and astrophysical observations.  These three avenues for study all contribute to our understanding of atoms and atomic processes, as well as to modelling of stellar atmospheres and their spectra.  However, they are not always exploited to their full potentials, and rarely coherently.  In particular, the use of high-quality stellar spectra in deriving basic atomic properties that can be observed rather directly in stellar spectra, such as the identification of atomic levels via measurement of spectral line wavelengths, has perhaps been underexploited.  Further, if the astrophysical modelling is sufficiently reliable, stellar spectra may be used to infer or place limits on properties directly related to the observed spectral line, such as the transition probability and perhaps the collisional broadening constant.  However, it must be realised that the inference of atomic properties only indirectly related to the observed spectral line, such as inelastic collision rates or interatomic interaction potentials, is fraught with danger as the problem is generally ill-posed.  To be authentic these properties must be derived from theory and experiment. Observations provide a test of the complete spectral modelling and thus an indirect check on the employed atomic data.

As demonstrated by this review, in recent decades, significant progress has been made on atomic physics relevant for the high accuracy interpretation of FGK stellar spectra.  A particular highlight has been the advancement in close-coupling type calculations for inelastic processes due to electron impacts.  Many experimental results and calculations derived with different methods\footnote{We note, however, that calculations usually lack an estimate of their uncertainty. This restricts their viability, because a number without an indication of its uncertainty has no real meaning.}, provides a reasonable picture of the uncertainties of the main electron collision processes in non-equilibrium modelling, at least for lighter atoms. Data for heavier atoms, especially the iron-group elements, are beginning to appear, but further calculations encompassing more atoms (especially neutrals) and excited states are needed.  Data for processes involving collisions of heavy particles has tended to trail behind those for electron collisions, for good reason.  But while several decades ago the outlook was bleak regarding calculations on the scale required for astrophysical modelling, now there is generally cause for optimism. Significant progress has been made on theoretical studies of collisions processes involving hydrogen, and more can be expected in the near future.  Of particular importance will be the extension of calculations of inelastic processes due to hydrogen atom collisions to complex atoms, such as the astrophysically important iron-group elements (e.g. Fe, Ti, Mn) and open-shell atoms (e.g. C, O) through both full-quantum calculations based on quantum chemistry calculations and approaches based on simplified models.  As regards broadening of spectral lines by hydrogen atom collisions, the next step in accuracy beyond the general methods currently in common use such as the LFU and ABO methods, which are not specific to any given atom at least in the case of neutrals, will almost certainly require calculations accounting for the detailed properties of the target atom.  Future calculations should consider the influence of the ionic configuration in both the interatomic potentials and the dynamics, ideally using quantum chemistry approaches, but also employing simplified approaches in order to enable calculations covering many atoms and transitions of astrophysical interest.  Extension of the calculation of self broadening of Balmer lines beyond H$\alpha$ using the unified line broadening theory would be of importance, ideally performed with detailed quantum chemistry calculations of the relevant potentials.

While there has been significant progress on theoretical calculations, experiments on these processes would be of great importance to future progress on accurate abundance analysis of FGK stars.  Such studies are needed both to guide theoretical calculations of these processes, as well as to help estimate uncertainties for cases where theory must be used for practical reasons related either to experimental challenges or limited resources.  Experiments on charge transfer involving hydrogen, particularly mutual neutralisation with the negative hydrogen ion, for lighter elements of astrophysical interest are expected in the coming years through ion-storage ring experiments, e.g. DESIREE.  Experiments on excitation by low-energy hydrogen atom impacts are difficult due to the problem of creating hydrogen atom beams at low eV energies \citep[see][]{Fleck1991}, and experiments on collisional broadening in shock tubes have been performed, but with rather low precision \citep[see][also \S~\ref{sec:metal}]{baird_width_1979,lemaire_broadening_1985}.  These attempts are many decades old, and new attempts at experiments on excitation, broadening and self broadening due to hydrogen atom collisions, with improved or novel experimental approaches, would be of significance for accurate studies of stellar atmospheres and their astrophysical applications, and are thus strongly encouraged.

The large quantities of atomic and molecular data needed for modelling stellar atmospheres and their spectra, means that compilations and databases such as NIST, Kurucz, VALD and VAMDC, play a crucial role at the interface between atomic and molecular physics and stellar astrophysics.  A critical issue in this respect is the credit received by the groups doing the theoretical and experimental work on which such databases, and by extension all modelling of stellar atmospheres and their spectra, are based.  In the present climate where bibliometrics play an important role in determining funding, this is an important problem to be solved if the current rate of progress is to be maintained, and perhaps even increased.

Finally, the propagation of uncertainties from atomic and molecular data through to astrophysical modelling is of great importance, and must be given more attention.  It provides not only an indication of the uncertainties in derived stellar properties, but quantitative feedback to atomic physics on where it is important to prioritise limited resources for the development of new theory and experimental methods, or the performing of calculations and experiments with existing methods.

\begin{acknowledgements}
This review owes particularly large debts to two great scientists, Jim O'Mara and Andrey Belyaev.  
I have benefitted immeasurably from having long-term scientific collaborations with them.  They have taught me not only physics and astrophysics, but been models of how to do science honestly and ethically.  I have learnt from them the value of analytic understanding, often with Jim on the ``back-of the-envelope''.  They showed me by example how to work carefully on hard problems on long timescales.  I am grateful to them both.

I thank the many other collaborators who have contributed to work presented here.  In particular, I wish to thank Stuart Anstee, Jenny Aspelund-Johansson, Juan Manuel Borrero, Remo Collet, Moncef Derouich, Alan Dickinson, Laine Falklund, Nicole Feautrier, Xavier Gad\'ea, Marie Guitou, Yeisson Osorio, Sylvie Sahal-Br\'echot, and Annie Spielfiedel, for collaboration on various problems in atomic physics over the years.  I thank also my many colleagues at Uppsala, past and present, for providing a stimulating and pleasant environment to work in.  I thank Igor Bray for clarifications regarding the CCC method.

This review and much of the work in it would not have been possible without financial support from the Royal Swedish Academy of Sciences, the Wenner-Gren Foundation, Göran Gustafssons Stiftelse and the Swedish Research Council.  For much of this work I was a Royal Swedish Academy of Sciences Research Fellow supported by a grant from the Knut and Alice Wallenberg Foundation. I am presently partially supported by the project grant ``The New Milky Way'' from the Knut and Alice Wallenberg Foundation.

\end{acknowledgements}

\bibliographystyle{spbasic}      
\bibliography{MyLibrary.bib,extra.bib}   

\end{document}